\documentclass[letter]{sheridan-template/sig-alternate-2013}

\newfont{\mycrnotice}{ptmr8t at 7pt}
\newfont{\myconfname}{ptmri8t at 7pt}
\let\crnotice\mycrnotice%
\let\confname\myconfname%

\permission{Permission to make digital or hard copies of all or part of this work for personal or classroom use is granted without fee provided that copies are not made or distributed for profit or commercial advantage and that copies bear this notice and the full citation on the first page. Copyrights for components of this work owned by others than ACM must be honored. Abstracting with credit is permitted. To copy otherwise, or republish, to post on servers or to redistribute to lists, requires prior specific permission and/or a fee. Request permissions from Permissions@acm.org.}
\conferenceinfo{CIKM'15,}{October 19--23, 2015, Melbourne, Australia.} 
\copyrightetc{\copyright~2015 ACM. ISBN \the\acmcopyr}
\crdata{978-1-4503-3794-6/15/10\ ...\$15.00.\\
DOI: http://dx.doi.org/10.1145/2806416.2806582}

\usepackage[
  pass,
]{geometry}
   
 \usepackage[utf8]{inputenc}
 \usepackage{pdfpages}
\usepackage{stfloats}
\usepackage{graphicx}
\usepackage{array}
\usepackage{amssymb}
\usepackage{amsmath}
\usepackage[ruled,vlined,linesnumbered,commentsnumbered]{algorithm2e}
\usepackage{caption}
\DeclareCaptionType{copyrightbox}
\usepackage{blindtext}
\usepackage{etoolbox}
\usepackage{subfig}
\usepackage[hyphens]{url}
\usepackage{balance}
\usepackage{mdwlist}
\usepackage{multirow}
\usepackage{placeins}
\usepackage{color}
\usepackage{changepage}
\usepackage{enumitem}

\usepackage[show]{chato-notes} 

\usepackage{booktabs}

\begin{document}

\def\myargmax{\mathop{\rm argmax}\nolimits}
\def\mysim{\mathop{\rm sim}\nolimits}
\def\myDCG{\mathop{\rm DCG}\nolimits}
\def\mynDCG{\mathop{\rm nDCG}\nolimits}

\title{Who With Whom And How? - Extracting Large Social Networks Using Search Engines}
 
\numberofauthors{1}
\author{
\alignauthor
Stefan Siersdorfer, Philipp Kemkes, Hanno Ackermann, Sergej Zerr
\\
\affaddr{L3S Research Center, Hannover, Germany}
\email{\{siersdorfer,kemkes,zerr\}@L3S.de}
\email{ackermann@tnt.uni-hannover.de}\\
}

\maketitle
\begin{abstract} 
Social network analysis is leveraged in a variety of applications such as
identifying influential entities, detecting communities with special interests, and determining the flow of information and innovations. However, existing approaches for extracting social networks from unstructured Web content do not scale well and are only feasible for small graphs.
In this paper, we introduce novel methodologies for query-based search engine
mining, enabling efficient extraction of social networks from large amounts of
Web data. To this end, we use
patterns in phrase queries for retrieving entity connections, and employ a
bootstrapping approach for iteratively expanding the pattern set. 
Our experimental evaluation in different domains demonstrates that our
algorithms provide high quality results and allow for scalable and efficient
construction of social graphs. 
\end{abstract}
\vspace{-5pt}

\vspace{-6pt}
\vspace{-2pt}

\section{Introduction}

Networking and communication are natural human activities that are massively supported through modern information technologies, most prominently the internet. Large networks of people are present in many parts of the Web, for instance, in form of contacts and friends in Social Web platforms, co-occurring named entities in web pages, or co-authors of articles in scientific portals. Social networks extracted from these information sources are exploited for various purposes: Node centrality measures in networks can help to identify important and influential persons or communities in areas such as entertainment, science, or politics. Information on social connections in customer networks can be leveraged as part of recommender systems that suggest items and new contacts to their users. Furthermore, knowing the topology of social graphs can shed light on the propagation of ideas and trust in social networks\mbox{~\cite{SocialInfluencePropagation, Anagnostopoulos:2008:ICS:1401890.1401897,Goyal:2010:LIP:1718487.1718518}}, and can enhance various IR applications such as personalized query expansion and media recommendation\mbox{~\cite{Bouadjenek:2013:SNS:2484028.2484131, Guy:2010:SMR:1835449.1835484}}. 

\begin{figure}[t]  
\center 

\includegraphics[width=0.44\textwidth]{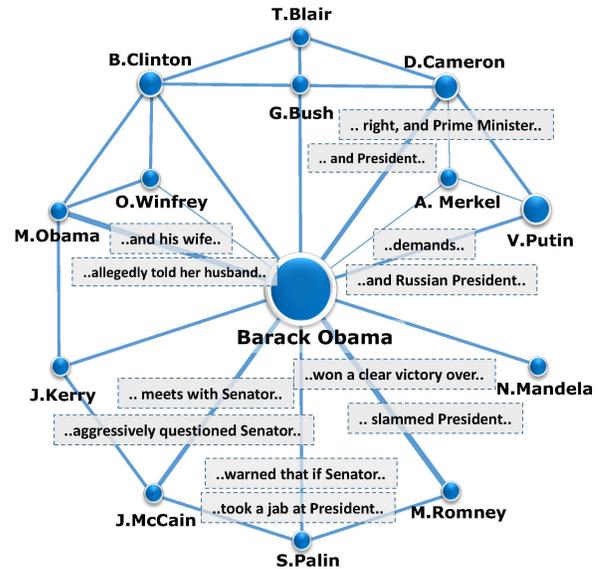}

   \vspace{-8pt} 
 \caption{Subgraph of a network extracted using our approach along with examples for connecting phrase patterns.}
 
 \label{fig:obama_graph} 
 \vspace{-20pt} 
\end{figure}
\begin{sloppypar}
In order to enable the analysis of social networks they have to be extracted from underlying data in the first place. Some sources already provide explicit and easy to extract information about user relations. This includes online platforms such as Facebook, Twitter, or LinkedIn that maintain user databases and offer software interfaces for accessing contacts, friends, or followers. Other sources - for instance, email corpora revealing communication links between persons, or scientific portals comprising information on co-authorship - contain more implicit, yet easy to extract network information. However, in many cases information about social connections is hidden within unstructured data such as Web pages, archives, and multimedia repositories. 
\end{sloppypar}

\begin{sloppypar} 
Search engines constitute our main access points to the Web, and there exists a number of works that employ search queries for detecting co-occurrences of entities and leverage these connections for extracting social networks~\cite{DBLP:journals/aim/KautzSS97,Matsuo:2006:PAS:1135777.1135837,nuray2009exploiting}. 

However, the suggested algorithmic approaches do not scale well and are only suitable for rather small datasets comprising just a few hundred or a few thousand nodes. 
\end{sloppypar}
\begin{sloppypar}
In this paper, we exploit pattern-based search engine queries for efficient
acquisition of large scale networks from unstructured Web data. These queries
consist of entities (e.g. ``Barack Obama'') and connecting patterns
(e.g. ``meets with'', ``and his wife'') leveraged for mining links between persons. 
Using such search patterns we can simultaneously explore a large number of connections with a single query. Starting with an initial set of entities, we iteratively expand both
entity pair and pattern sets. We identify connections which are
likely to be important and use this information to formulate subsequent queries,
thereby greatly reducing the number of search engine requests.
For illustration, a small excerpt of a constructed graph is shown in Figure~\ref{fig:obama_graph}.
\end{sloppypar}

\vspace{7pt}
The main contributions of this work can be summarized as follows: 
\vspace{5pt}
\begin{itemize}[leftmargin=6mm]
\item We introduce intelligent prioritization approaches for efficiently expanding social networks using pattern based search engine queries.
\item We employ a bootstrapping approach for covering multiple aspects of social relations by iteratively extending search pattern and entity sets for discovering domain specific social connections.
\item We conduct an extensive experimental evaluation demonstrating the high efficiency and accuracy of our methods clearly outperforming the baseline. Furthermore, we analyze various aspects of the extracted social graphs, such as their structural properties, connecting patterns, and strength of social relations. 
\end{itemize}

The remainder of this paper is organized as follows:
Section~\ref{sec:Related_Work} discusses related work on social network construction from semi-structured data sources, search engine based network mining, and extraction of semantic relationships. 
In Section~\ref{sec:methods} we describe our methods for network extraction including search patterns and their expansion based on bootstrapping as well as different search prioritization criteria. The evaluation presented in Section~\ref{sec:evaluation} studies the cost efficiency and accuracy of our methods, and provides further insights about the structure of the extracted networks. 
Finally, in Section~\ref{sec:Conclusions_and_Future_Work} we conclude and
describe directions of our future work. 


\vspace{10pt}
\section{Related Work}
\label{sec:Related_Work}
\vspace{10pt}

There is a plethora of work on social network extraction from text and visual data. In~\cite{Elson:2010:ESN:1858681.1858696}, for instance, social connections between fictional characters are inferred from dialogs in books, and, similarly, in~\cite{ottoman} a social network is extracted from the narrative of an Ottoman scholar and world traveler. In~\cite{DBLP:conf/socinfo/WienekeDSLCLNPFTMNMHM13} social connections are constructed from a historical multimedia repository by leveraging co-occurrences of faces in images. Bird et al.~\cite{Bird:2006:MES:1137983.1138016} extract social networks using information about senders and recipients obtained from headers in email corpora. Social networks extracted from other (semi-)structured sources include academic networks or co-citation graphs in publication databases~\cite{Mohaisen:2010:MMT:1879141.1879191,Tang:2008:AEM:1401890.1402008}.

Information extraction tackles the problem of deriving structured information from unstructured data. 
There is a body of work on semantic relationship mining based on textual patterns~\cite{bach2007review,konstant2014}. 
In~\cite{DBLP:conf/acl/PantelP06} a bootstrapping algorithm is employed for iteratively discovering new patterns and semantic connections. In~\cite{Pasca:2004:ACN:1031171.1031194} and~\cite{Cafarella:2009:WES:1519103.1519112} categorical facts and concept hierarchies are extracted from web and news corpora as entity-pattern-entity triples. Cimiano et al.~\cite{Cimiano:2004:TSW:988672.988735,Cimiano:2005:GCC:1060745.1060796} leverage pattern-based search engine queries for building ontologies. In~\cite{Jain:2009:ICE:1645953.1646198} comparable entities such as consumer items are extracted using an expandable set of contrasting expressions (e.g. ``vs.'' and ``instead of''); Jiang et al.~\cite{Jiang:2013:LOC:2541176.2505666} use similar relations for constructing entity graphs and product recommendations in web search.
The KnowItAll system~\cite{knowitall} employs a supervised learning approach for validating facts extracted using search engine queries based on manually provided pattern templates. Open information extraction systems such as TextRunner~\cite{Yates:2007:TOI:1614164.1614177} employ shallow parsing, and, as patterns, employ normalized expressions between noun phrases in an unsupervised manner. The output of these systems are databases that contain fact tuples such as ``{\em (Berkley, located in, Bay Area)}'' or ``{\em (Oppenheimer, professor of, theoretical physics)}''. 
In contrast to works focusing on general fact mining, we provide a novel methodology for extracting social networks. Our contributions include the systematic construction of connected social graphs, their efficient and intelligent expansion, as well as the analysis of their structural properties.

The first approach employing search engines for network mining is probably the work by Kautz et al.~\cite{DBLP:journals/aim/KautzSS97}. The authors exploit co-occurrences of names on result pages to
identify connections between persons; the resulting networks are
relatively small and just centered around a single person. The POLYPHONET
system~\cite{Matsuo:2006:PAS:1135777.1135837,DBLP:conf/aaai/MatsuoTN07}
determines co-occurrence counts in Google queries to identify pairwise
connections. In addition, different types of relationships and correlations
with topic-related keywords are taken into
account to disambiguate entities. In~\cite{Canaleta:2008:SES:1566899.1566939} search engines are used
to retrieve connected email addresses. The Flink
system~\cite{Mika:2005:FSW:1741305.1741326} focuses on the mining of author
networks and combines web search with information from emails and
publications. In~\cite{DBLP:conf/rskt/NasutionN10} co-occurrences of persons
in search result snippets are leveraged. There are a few attempts to increase
efficiency in discovering entity relations on the web~\cite{he2007efficient,
 nuray2009exploiting}; they focus on snippet clustering and
entity disambiguation to reduce the number of search requests. However, all of the current methods are based on pair-wise entity
queries, making them infeasible for mining larger graphs.
\vspace{-4pt}
\\\\ 
To the best of our knowledge, our methods are the first to go beyond pair-wise querying for social connections and use intelligent query pattern construction and prioritization approaches for graph expansion that allow for tunable and larger scale network construction. 
 

\section{Social Graph Mining}
\label{sec:methods}

In this section we describe our methodology for mining social graphs using search engines. To this end, we first introduce an efficient basic algorithm using web search with a fixed pattern set for identifying connections between persons. We then extend this method by an iterative approach for automatically collecting new patterns. Finally, in order to further improve the cost efficiency of network construction, we introduce methods for prioritizing nodes for graph expansion. 
\subsection{Mining Graphs with a Static Set of Query Patterns}
\label{sec:bfs}
\vspace{3pt}
\begin{sloppypar}
Given an initial seed set of entities \textit{I} and a set of patterns
\textit{P}, we use connectivity search queries of the form \mbox{{\em <entity> <p>}} (where
{\em <p>} is a connecting pattern as e.g. in ``Barack Obama meets'') in order to identify links to other entities. We iteratively issue these queries over an initial seed set of entities (e.g. a list of a politicians). In this way we obtain links between entities in the seed set as well as new entities. This is repeated for all patterns in our pattern set (in our experiments we used a small initial set consisting of the single term ``and''). For snippets of the top-$k$ search results in Bing we check for exact strings of the form \mbox{{\em <entity> <p> <entity>}} where both entities match a person name in a database (we were using entities from DBpedia and IMDb in our experiments).\footnote{\small We decided to keep the entity recognition sub-component simple and lightweight. More enhanced NLP and disambiguation techniques are out of the scope of this work, but could help to further improve our already very good results.} The process of querying entities is repeated for the newly discovered entities in a ``breadth-first'' manner. The resulting (undirected) edges between two entities are weighted by the overall occurrence count in the snippets across all patterns, omitting edges whose weight is below a threshold $\tau$.
\end{sloppypar}
Algorithm~\ref{alg:static-patterns} shows the details of the algorithm. The network and a priority queue of nodes are initialized based on the entity seed set (lines 2--4). Here the priority queue is a simple ``first-in-first-out'' queue; in Section~\ref{sec:prioritization} we will introduce more enhanced mechanisms for prioritization. The nodes in the queue serve as input for pattern-based queries (line 8), and query results are used for expanding the social network and updating the queue (lines 9--17). The network expansion continues until the query budget is spent or until there are no unvisited nodes left (line 19). The running time of the algorithm is linear in the sum of nodes and edges visited and found.

\subsection{Iterative Pattern Mining}
\label{sec:Iterative}

We extend the Graph Mining method described above through an approach for
iteratively discovering new patterns. To this end, we start with an initial set of
seed patterns and extend it in each round of the algorithm, taking into account the weights of edges 
extracted in that round. 
\begin{sloppypar}
We then iterate over the top-$h$
resulting edges with the highest weight and use them to issue queries of the
form {\em ``<entity>'' ``<entity>''}. Snippets of the search results are
then checked for strings of the form \mbox{{\em <entity> <c> <entity>}},
where {\em <c>} is an arbitrary string between two different entities; {\em <c>} becomes a
candidate for a new pattern. Among the candidates we want to identify patterns
with high coverage in the result snippets. To this end, let $n$ be number of
occurrences of candidate pattern {\em <c>} (support), $m$ be the number of distinct
entity pairs (diversity) and $d$ be the number of
distinct domains (spam resistance) comprising the pattern.
We include {\em <c>} into the pattern set if $n \cdot m \cdot
d^2$ is above a threshold $\sigma$. Extracting top-k patterns can be managed
effectively using a priority queue with $log(n)$ running time for
operations like insertion and updating.
This pattern mining method is employed after each complete iteration of the
algorithm described in the previous paragraph.
\end{sloppypar}
\begin{algorithm}[t*]
\KwIn{
$I$: initial set of entities \\
$IP$: initial set of patterns \\
$\tau$: threshold for edge weights\\
$\sigma$: threshold for pattern weights\\
$maxIter$: maximum number of iterations\\
}
\KwOut{social graph $G = (V,E)$}
\Begin
{
	$E \leftarrow \emptyset$ \\
	$V \leftarrow I$ \\
	$Cand \leftarrow I$ \tcp{\small graph extension candidates}
	$P \leftarrow IP$ \tcp{\small pattern set}
	$iter \leftarrow 0$ \\
	\Repeat{$iter = maxIter$}
	{
		$NewCand \leftarrow \emptyset$\\
		$NewPatterns \leftarrow \emptyset$\\
		\For{$e \in Cand$}
		{
			$edges \leftarrow search(e,P,\tau)$ \tcp{\small web search for edges with node $e$, using query patterns $P$ and weight threshold $\tau$}
			\For{$(\{e,e_i\},w) \in edges$}
			{
				\If {$\{e,e_i\} \not \in E$}
				{
					$V \leftarrow V \cup \{e_i\}$ \\
					$NewCand \leftarrow NewCand \cup \{e_i\}$ \\ 
					$E \leftarrow E \cup \{(\{e,e_i\},w)\}$ \\ 
				}
				\Else
				{
					$E \leftarrow E \setminus \{(\{e,e_i\},x)\}$ \\
					$E \leftarrow E \cup \{(\{e,e_i\},x+w)\}$ \\
				}
			}
		}
		$Cand \leftarrow Cand \cup NewCand$ \\
		$NewPatterns \leftarrow pSearch(E,\sigma)$ \tcp{\small web search for patterns, using edges from $E$, and weight threshold $\sigma$}
		$P \leftarrow P \cup NewPatterns$; 
		$iter \leftarrow iter+1$ \\
	}
	\Return{$(V,E)$}
		
}
\caption{Graph Construction with Pattern Set Expansion}
\label{alg:expanded-patterns}
\vspace{10pt}
\end{algorithm} 
The details of the algorithm are shown in
Algorithm~\ref{alg:expanded-patterns}. Seed patterns and entities are first used
to initialize the bootstrapping approach (lines 2--5). Then, for a fixed number
of iterations, the social network is expanded based on the current query
patterns (lines 11--19), and, conversely, the pattern set is expanded using the
current edges in the network (lines 21--22). The running time of the algorithm
is linear in the sum of nodes and edges visited, multiplied by the number of
patterns found.

\subsection{Prioritization of Nodes for Network Expansion}
\label{sec:prioritization} 

The algorithms described so far expand the social graph in a ``breath-first-search'' manner. The advantage of this approach is its good coverage: Seedset and subsequentially found nodes are systematically visited and expanded until our budget of connectivity queries is exhausted. However, in order to further optimize budget usage we might want to put higher priority on more ``essential'' nodes during the network expansion process. In the following we focus on two criteria for prioritizing nodes: popularity and novelty. 

Popularity in this context intuitively corresponds to the importance of nodes within the network. We employ the number of inlinks $\rho$ in the current version of the social graph during the extraction process as popularity measure. Although very effective in our experiments, one could easily imagine using alternative node centrality measures such as PageRank~\cite{ilprints422}, hub scores in HITS~\cite{Kleinberg:1999:HAC:345966.345982}, or one of their various variants (e.g.~\cite{Langville:2006:GPB:1146372, Jin:2010:TID:1772690.1772740}). 
\begin{algorithm}[!t]
\KwIn{
$I$: initial set of entities \\
$P$: set of patterns \\
$\tau$: threshold for edge weights\\
$maxReq$: maximum number of requests\\
$pQueue$: A container which sorts the candidates according to the employed algorithm
}
\KwOut{social graph $G = (V,E)$}
\Begin
{
	$E \leftarrow \emptyset$ \tcp{\small Edges}
	$V \leftarrow I$ \tcp{\small Vertices}	
	$pQueue.addAll(I)$ \tcp{\small graph extension candidates}
	$requestCnt \leftarrow 0$ \\

	\Repeat{$requestCnt \ge maxReq$ or $pQueue.empty()$}
	{	 
		$e \leftarrow pQueue.pop()$
	
		$(edges, requests) \leftarrow search(e,P,\tau)$ \tcp{\small web search for edges with node $e$, using query patterns $P$ and weight threshold $\tau$}
		\For{$(\{e,e_i\},w) \in edges$}
		{
			\If {$\{e,e_i\} \not \in E$}
			{
				$V \leftarrow V \cup \{e_i\}$ \\				
				$E \leftarrow E \cup \{(\{e,e_i\},w)\}$ \\ 
				\If {$e_i \not \in pQueue$}
				{
					$pQueue.push(e_i)$\\
				}
			}
			\Else
			{
				$E \leftarrow E \setminus \{(\{e,e_i\},x)\}$ \\
				$E \leftarrow E \cup \{(\{e,e_i\},x+w)\}$ \\
			}
		}
	
		$requestCnt \leftarrow requestCnt+requests$ \\
	}
	\Return{$(V,E)$}
	
}

\caption{Graph Construction with Static Pattern Set}

\label{alg:static-patterns}
\vspace{10pt}
\end{algorithm} 

\vspace{6pt}
Novelty, on the other hand, corresponds to how recently a node was added in the course of our network construction process. Intuitively, exploring more recent nodes can help to reach out more quickly to new communities within the network. We combine popularity and novelty into a priority measure $\Phi$ for a node $v$ using exponential temporal decay as follows: $\Phi(v) = \rho \cdot e^{-\alpha \cdot t}$,
where our time measure $t$ corresponds to the number of expansion node steps conducted since the node was added to the queue, and $\alpha$ is a parameter for balancing the influence of popularity and novelty. 
Exponential decay is often used for time dependent ranking of items
in the context of IR (see e.g. TempPageRank~\cite{Yu:2004:TDS:1013367.1013519}).
This approach can be directly integrated in the framework provided in Algorithm~\ref{alg:static-patterns} using a priority queue based on the $\Phi$ score, and could, in principle, also be integrated into the dynamic pattern mining process (Algorithm~\ref{alg:expanded-patterns}).

\vspace{50pt}
\section{Evaluation}
\label{sec:evaluation}

In this section we evaluate the network extraction methods described in Section~\ref{sec:methods}, starting with entities from two domains: politics and movies. The objective of our evaluation was to study (1) the cost efficiency of different strategies in terms of network size obtained with a given budget of search engine requests, (2) the accuracy of the extracted networks, and (3) the structure of the social graphs as well as properties of discovered social relations. 
In our experiments, more than 300,000 persons and 1.4 million social connections were retrieved using two seed sets containing just 342 politicians and 100 actors. The resulting networks are coined WikiNet and IMDbNet.


\subsection{Setup}
\label{sec:graph_generation_setup}

\paragraph*{Domains and Seed Sets} For initializing our algorithms we used two entity seed sets from political and movie domains as follows: 

\begin{itemize}

\item {\em WikiNet seed}: We extracted 342 politician names from the Wikipedia list of the current heads of state and government from all over the world\footnote{\small \url{http://en.wikipedia.org/wiki/List_of_current_heads_of_ state_and_government}}. 

\item {\em IMDbNet seed}: We considered the 100 current and former leading actors as listed in IMDB\footnote{\small \url{http://www.imdb.com/list/ls050274118/}}. 

\end{itemize}
In order to simplify named entity recognition in search engine snippets, we restricted the set of possible persons in our WikiNet to those listed in \textit{DBpedia Version 2014} (containing about 1 million distinct person names) and in the IMDbNet to names from the IMDb directory (about 2 million distinct names).

\paragraph*{Implementation Details} We employed the Bing Search Engine API for issuing our network expansion queries. Preliminary tests showed that in most cases the overall result list returned per query did not exceed $k=200$ entries, although the estimated number of results often went beyond thousands. Therefore, we issued 4 requests per query as Bing provides 50 results per request (if available). In order to avoid recurring API requests we simultaneously cached results from previous queries. The search API has some technical limitations, including the fact that a few (special) characters, like ``\&'', ``,'', ``+'' cannot be used as search query. Furthermore, most of the snippets returned for disjunctive queries do not contain query terms, and, thus, cannot be used for relation extraction.

\paragraph*{Tested Strategies} 
We evaluated the following methods: 
\begin{itemize}

\item The breadth-first graph mining approach with fixed pattern set described in Section~\ref{sec:bfs} ({\bf BF}).
 
\item The prioritization approach from Section~\ref{sec:prioritization}, with different values for the tunable decay parameter $\alpha$ (0, 0.005, and 0.01) ({\bf Prio}). 

\item The iterative pattern mining approach described in Section~\ref{sec:Iterative} ({\bf PatternIter}).

\end{itemize}
We set the edge weight threshold $\tau$ to 2 for all algorithms. 
For {\em PatternIter} we set the cut-off value $\sigma$ for pattern selection to 5, and used value $h=100$ for 
top connections employed for discovering new patterns. 
The initial query pattern sets used in our experiments consisted only of the
general pattern ``and''; for extracting connections from snippets we employed an
additional small set of manually selected patterns\footnote{\small The following patterns were used:
``meets'', ``\textvisiblespace'', ``\&", ``,'', ``speaks with'', ``und'', ``et'', ``y'', ``-''}.
\vspace{-5pt}
\paragraph*{Baseline} All of the current works on social graph construction using search engines are based on pairwise entity queries. As {\bf baseline} we tested the method described in~\cite{Matsuo:2006:PAS:1135777.1135837} (POLYPHONET) which tries to reduce the number of pair-wise queries by identifying subsets of promising query pairs. More specifically, starting with a seed set of entities, for each entity a list of candidate entities (contained in the entity set) is extracted from Web pages obtained by querying for the entity. Then queries for the pairs of seed entities and the candidates for associated entities are issued in order to obtain co-occurrence values based on search result counts, and edges with a value above a threshold $t$ are included in the graph. 
In order to allow for network expansion beyond the seed set, we included further entities found in the Web pages for identifying additional query candidates.

\begin{table}[t]
\center
\def\arraystretch{1.15}%
\begin{tabular}{llrr}
\cline{2-4}
\multicolumn{1}{c|}{}           & \multicolumn{1}{l|}{\bf Algorithm}  & \multicolumn{1}{c}{\bf Nodes} & \multicolumn{1}{c|}{\bf Edges}    \\ 
\cline{2-4} \vspace{-3.5mm} \\
\hline\multicolumn{1}{|l|}{\multirow{6}{*}{WikiNet}}             & \multicolumn{1}{l|}{baseline $t$=0.1} & 6,956            & \multicolumn{1}{r|}{13,571} \\
\multicolumn{1}{|l|}{}             & \multicolumn{1}{l|}{baseline $t$=0.4} & 2,925            & \multicolumn{1}{r|}{3,838}  \\ \cline{2-4} 
\multicolumn{1}{|l|}{} & \multicolumn{1}{l|}{BF}     & 113,988          & \multicolumn{1}{r|}{368,806}    \\
\multicolumn{1}{|l|}{}              & \multicolumn{1}{l|}{decay $\alpha$=0}  & 98,479          & \multicolumn{1}{r|}{456,223}    \\
\multicolumn{1}{|l|}{}              & \multicolumn{1}{l|}{decay $\alpha$=0.005} & 110,260          & \multicolumn{1}{r|}{346,663}    \\
\multicolumn{1}{|l|}{}              & \multicolumn{1}{l|}{decay $\alpha$=0.01} & 116,081          & \multicolumn{1}{r|}{323,372}    \\ \hline \hline

\multicolumn{1}{|l|}{\multirow{6}{*}{IMDbNet}}             & \multicolumn{1}{l|}{baseline $t$=0.1} & 7,159            & \multicolumn{1}{r|}{34,903} \\
\multicolumn{1}{|l|}{}             & \multicolumn{1}{l|}{baseline $t$=0.4} & 2,192            & \multicolumn{1}{r|}{6,666}  \\ \cline{2-4} 

\multicolumn{1}{|l|}{}              & \multicolumn{1}{l|}{BF}     & 109,453          & \multicolumn{1}{r|}{376,429}    \\
\multicolumn{1}{|l|}{}              & \multicolumn{1}{l|}{decay $\alpha$=0}  & 107,085          & \multicolumn{1}{r|}{425,830}    \\
\multicolumn{1}{|l|}{}              & \multicolumn{1}{l|}{decay $\alpha$=0.005}  & 112,736          & \multicolumn{1}{r|}{264,679}    \\
\multicolumn{1}{|l|}{}              & \multicolumn{1}{l|}{decay $\alpha$=0.01} & 112,782          & \multicolumn{1}{r|}{242,184}    \\ \hline
\end{tabular}
\vspace{-5pt}
\caption{Number of nodes and edges obtained for different strategies.} 
\label{tab:cost_algorithms_overall}
\vspace{-15pt}
\end{table}

\begin{table*}[t]
\center
\def\arraystretch{1.2}%
\begin{tabular}{llrrcrrc}
\cline{3-8}
\multicolumn{1}{c}{}               & \multicolumn{1}{c|}{}        & \multicolumn{3}{c|}{\bf Node Degree}                             & \multicolumn{3}{c|}{\bf Edge Weight}  \\                          
 \cline{2-8} 
\multicolumn{1}{c|}{\bf }           & \multicolumn{1}{l|}{\bf Algorithm}   & \multicolumn{1}{c}{\bf Mean} & \multicolumn{1}{c}{\bf SD} & \multicolumn{1}{c|}{\bf Median} & \multicolumn{1}{c}{\bf Mean} & \multicolumn{1}{c}{\bf SD} & \multicolumn{1}{c|}{\bf Median} \\ 
\cline{2-8} \vspace{-3.7mm}\\
\hline
\multicolumn{1}{|l|}{\multirow{4}{*}{WikiNet}} & \multicolumn{1}{l|}{BF}      & 6.47           & 12.53                 & \multicolumn{1}{c|}{2}   & 8.53           & 23.62            & \multicolumn{1}{c|}{3}   \\
\multicolumn{1}{|l|}{}              & \multicolumn{1}{l|}{decay $\alpha$=0}   & 9.27           & 16.24                 & \multicolumn{1}{c|}{2}   & 8.84           & 23.93            & \multicolumn{1}{c|}{3}   \\
\multicolumn{1}{|l|}{}              & \multicolumn{1}{l|}{decay $\alpha$=0.005}   & 6.29          & 9.46                 & \multicolumn{1}{c|}{2}   & 7.65          & 18.43            & \multicolumn{1}{c|}{3}   \\
\multicolumn{1}{|l|}{}              & \multicolumn{1}{l|}{decay $\alpha$=0.01} & 5.57           & 8.39                  & \multicolumn{1}{c|}{2}   & 7.65           & 17.95            & \multicolumn{1}{c|}{3}   \\ \hline \hline
\multicolumn{1}{|l|}{\multirow{4}{*}{IMDbNet}}   & \multicolumn{1}{l|}{BF}      & 6.88           & 14.03                 & \multicolumn{1}{c|}{1}   & 8.69           & 24.11            & \multicolumn{1}{c|}{3}   \\
\multicolumn{1}{|l|}{}              & \multicolumn{1}{l|}{decay $\alpha$=0}   & 7.95           & 15.43                 & \multicolumn{1}{c|}{1}   & 8.97           & 25.04            & \multicolumn{1}{c|}{3}   \\
\multicolumn{1}{|l|}{}              & \multicolumn{1}{l|}{decay $\alpha$=0.005}   & 4.70          & 7.60                 & \multicolumn{1}{c|}{2}   & 7.43          & 19.14            & \multicolumn{1}{c|}{3}   \\
\multicolumn{1}{|l|}{}              & \multicolumn{1}{l|}{decay $\alpha$=0.01} & 4.29           & 6.74                  & \multicolumn{1}{c|}{2}   & 7.35           & 18.67            & \multicolumn{1}{c|}{3}   \\ \hline
\end{tabular}
\vspace{-6pt}
\caption{Summary statistics for node degrees and edge weights in the extracted networks.} 
\label{tab:summary_stats}
\vspace{-6pt}
\end{table*}
\subsection{Cost Efficiency}
\label{sec:graph_generation_cost efficiency}

Table~\ref{tab:cost_algorithms_overall} shows the number of nodes and edges in the expanded networks computed by our methods using the same seed sets and exhausting a fixed budget of 200k queries per run. The main observations are the following:
\begin{itemize} 
\vspace{-5pt}
\item Starting with the small seed sets consisting of just a few hundred
entities described in the setup, our methods are able to extract networks
containing around 100,000 entities, and between 200,000 and 450,000 edges for
both domains. Our quality oriented evaluation in Section~\ref{sec:evaluation} will
demonstrate the high accuracy of edges generated by our methods.
\vspace{-5pt}
\item The tunable decay parameter $\alpha$ in the prioritization approach {\em Prio} is applicable for trading novelty against popularity. This is reflected in both domains by the larger number of social relations extracted for lower values of $\alpha$ and the larger number of nodes found for higher values of $\alpha$. 

\item Our new methods clearly outperform the {\em baseline} which does not scale well for expanding entity sets and produces much smaller networks for the given budget. Even for a small co-occurrence threshold $t=0.1$ the networks obtained by the baseline method are already between one and two orders of magnitude smaller. For $t=0.4$ (suggested in POLYPHONET for extracting relations of acceptable quality) the reduction in network growth becomes even more apparent.

\end{itemize}

Note, that in addition to the number of search engine requests taken into account for this evaluation, the baseline also introduces a substantial overhead by requiring the download and processing of a large set of Web pages (about 85,000 pages in our experiments). In contrast, our methods avoid these extra costs by working directly on search result snippets. 

\begin{figure*}[t] 
\center 
\subfloat[{WikiNet edges}]{\label{fig:poli_imdb_edges} 
\includegraphics[width=0.32\textwidth]{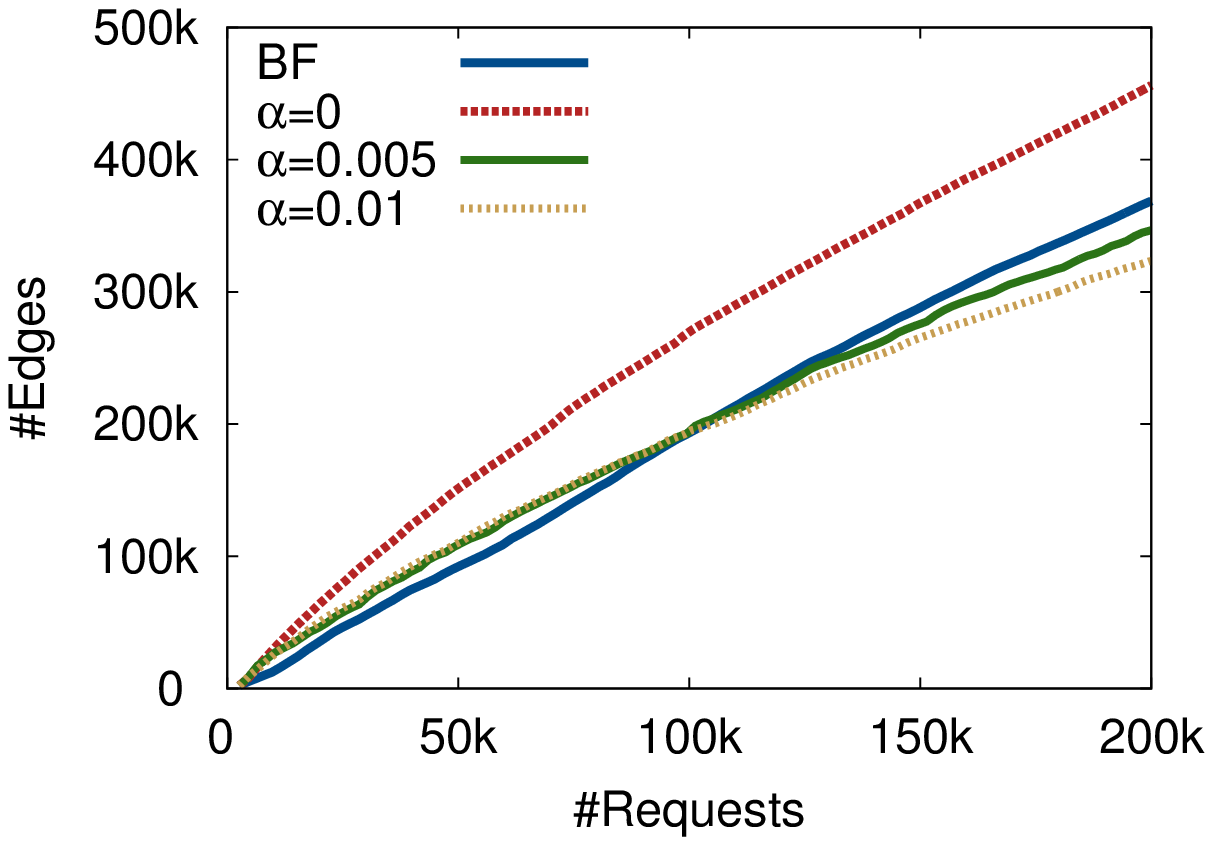}}
\subfloat[{WikiNet nodes}]{\label{fig:poli_imdb_nodes} 
\includegraphics[width=0.32\textwidth]{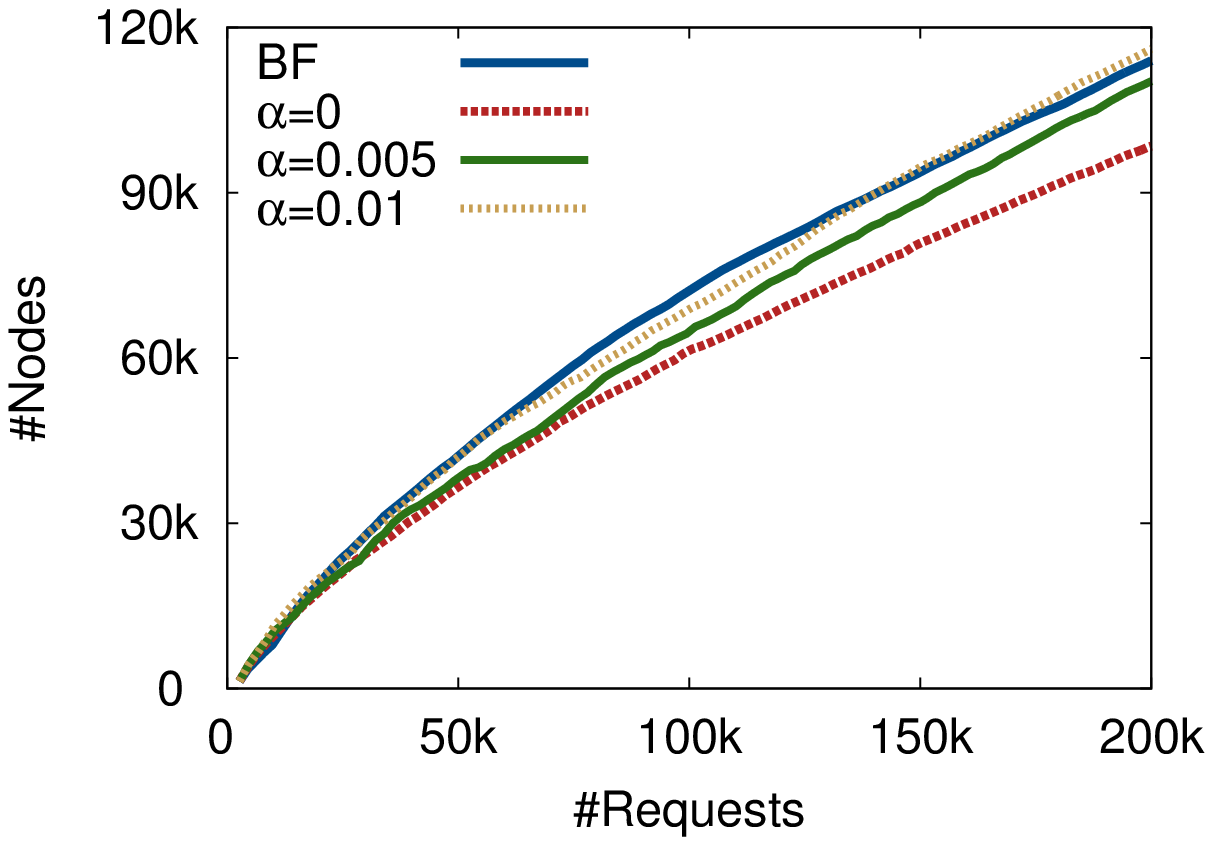}}
\vspace{-6pt}
\\
\subfloat[{IMDbNet edges}]{\label{fig:cost_imdb_edges} 
\includegraphics[width=0.32\textwidth]{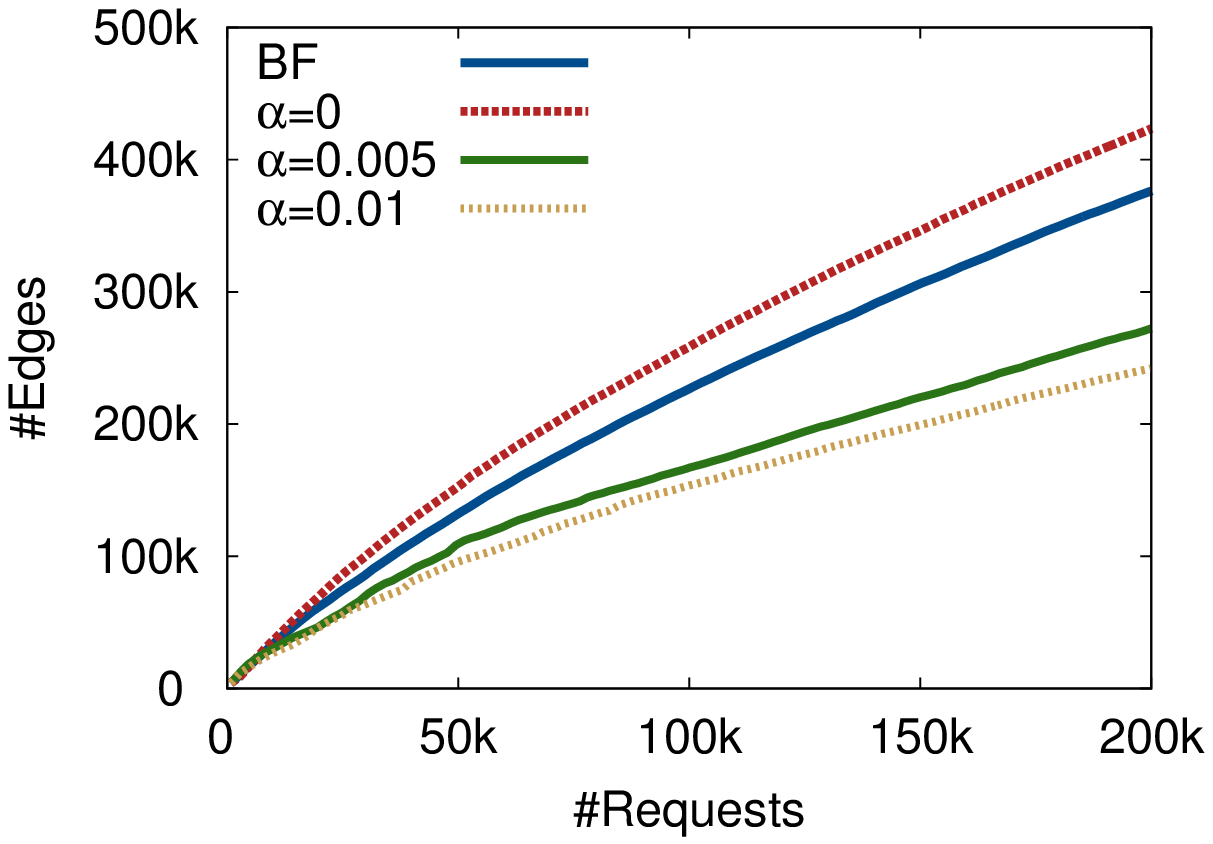}}
\subfloat[{IMDbNet nodes}]{\label{fig:cost_imdb_nodes} 
\includegraphics[width=0.32\textwidth]{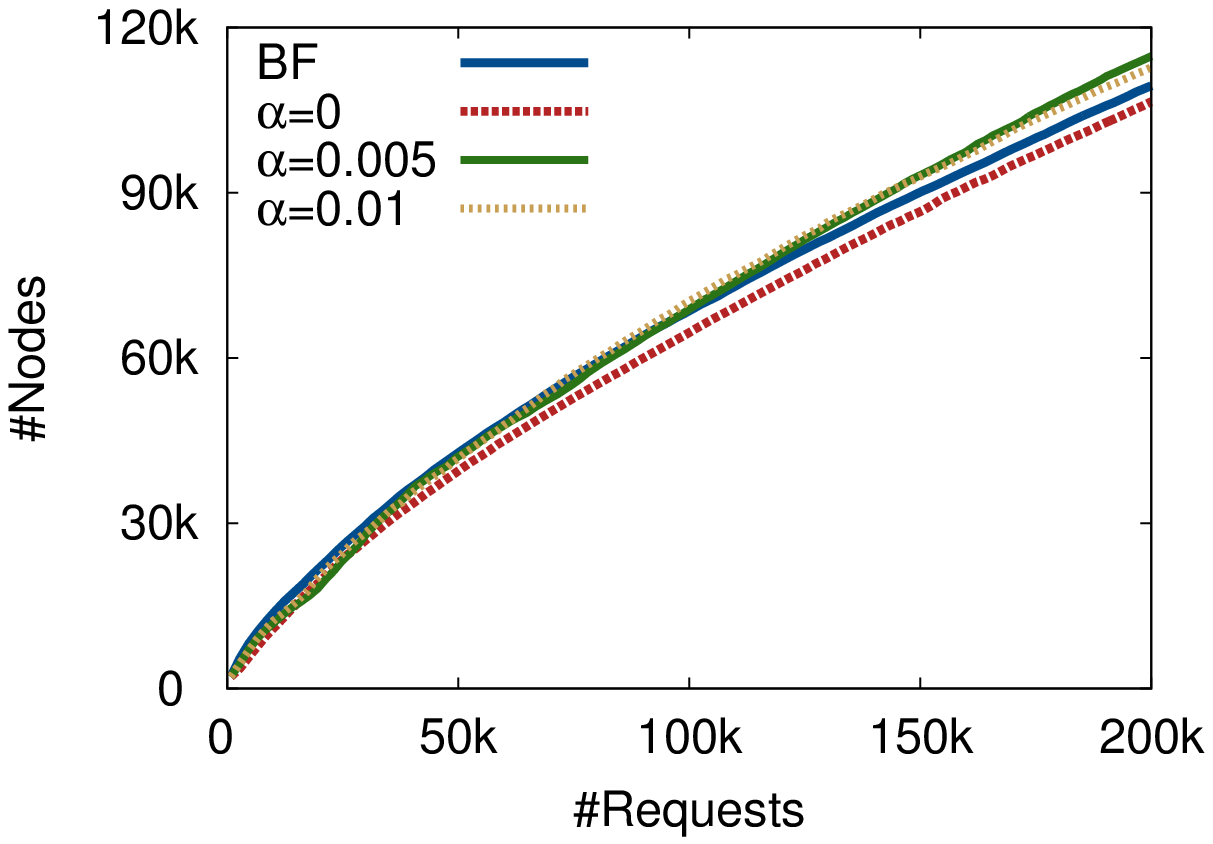}}
\vspace{-6pt}
\caption{ Nodes and edges versus requests spent for WikiNet and IMDbNet.} 
\label{fig:cost_algorithms}
\vspace{-6pt}
\end{figure*}

Figure~\ref{fig:cost_algorithms} provides additional details about the network size development with respect to the amount of search engine requests spent. The effect of a stronger emphasis on novelty (corresponding to higher $\alpha$) is reflected in a slight increase in the number of new nodes, starting from around 50k -- 100k requests. On the other hand, prioritization of popular nodes with high in-degree (lower $\alpha$) leads to a high increase in the number of edges (starting from around 10k requests for both datasets).

\vspace{-2pt}
\subsection{Network Properties}
In the following we describe structural properties of the extracted networks as well as the development of the networks during the execution of our algorithms.

\begin{figure}[t]
\vspace{-7pt} 
 \center 
\includegraphics[width=0.32\textwidth]{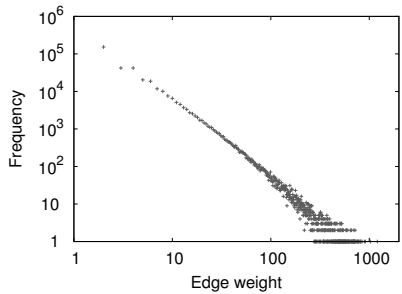}
\vspace{-6pt}
\caption{ Edge weight distribution for the {\em BF} algorithm applied on
WikiNet seed set\\}
\vspace{-12pt}
\label{fig:edge-weights}
\end{figure}

\paragraph*{Node Degree and Edge Weight Distributions}
Table~\ref{tab:summary_stats} provides summary statistics for the distribution of nodes and edges in the extracted networks. For all experiments we observe a right-skewed distribution of nodes and edges with median node degree of 1 or 2 and median edge weight of 3. (The edge weights in all the experiments follow a power law distribution as exemplarily depicted in Figure~\ref{fig:edge-weights} for the {\em BF}-constructed WikiNet.) The decay parameter $\alpha$ for the \emph{Prio} strategy has a clear impact on the distributions: Decreasing the value of $\alpha$ results in an increase of the means for both node degree and edge weights due to the higher prioritization of nodes with already high degree, and also results in a wider ``spread'' of the values as reflected by a higher standard deviation. 
 
\begin{figure*}[!hbp] 
 \center 
\subfloat[{WikiNet BF}]{\label{fig:nd_wikipedia_0}
\includegraphics[width=0.3\textwidth]{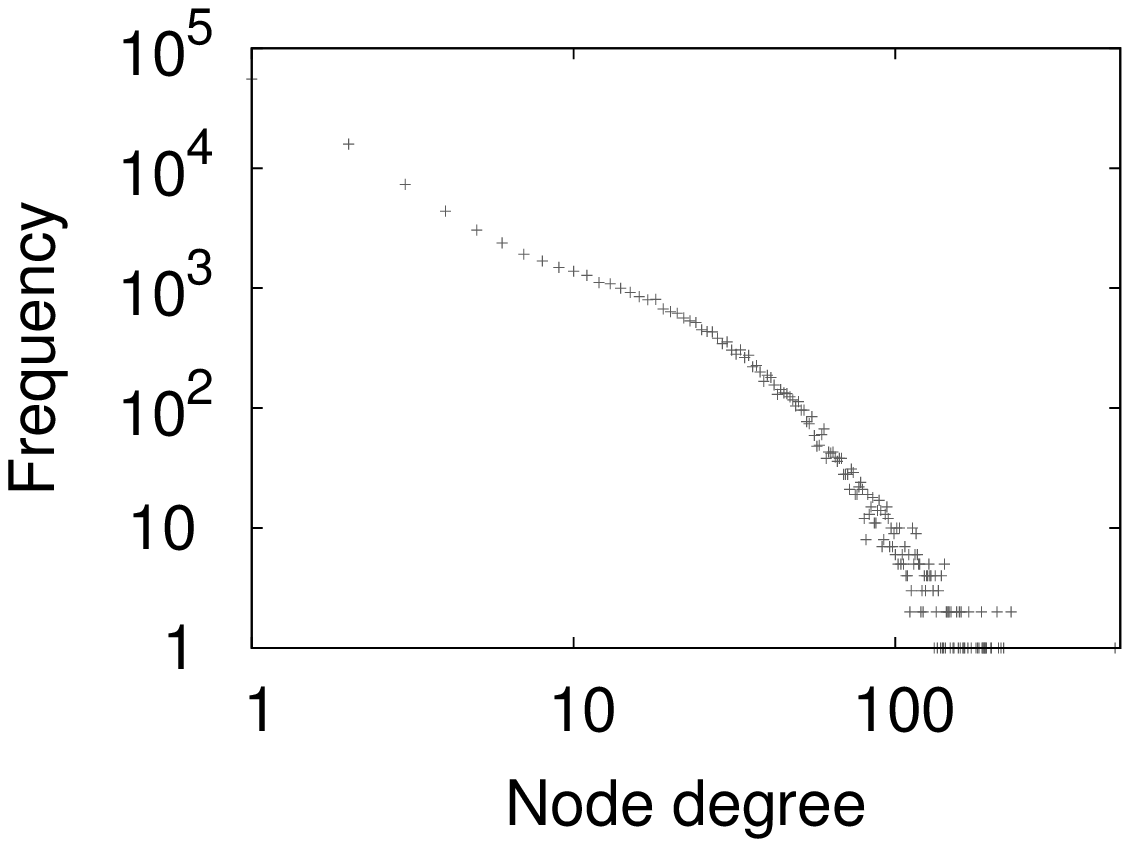}}
\subfloat[{WikiNet $\alpha$=0.01}]{\label{fig:nd_wikipedia_1}
\includegraphics[width=0.3\textwidth]{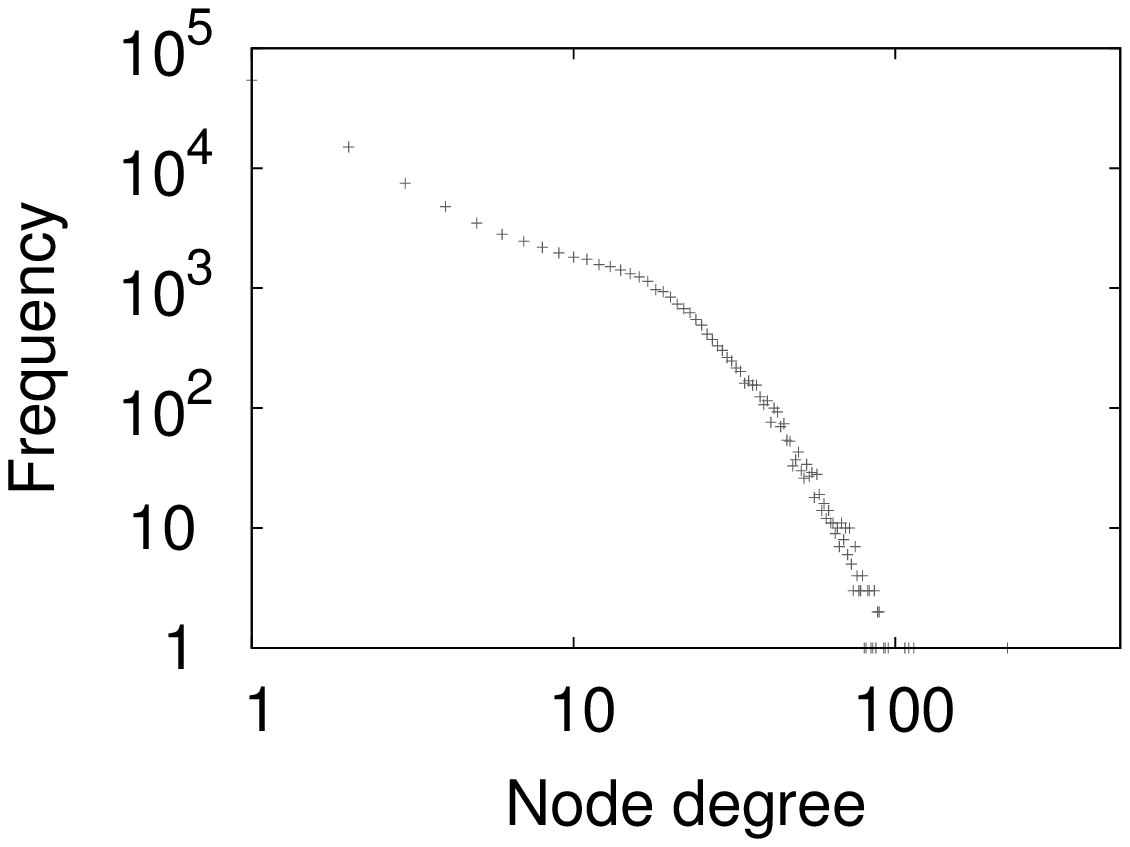}}
\subfloat[{WikiNet $\alpha$=0.0}]{\label{fig:nd_wikipedia_2}
\includegraphics[width=0.3\textwidth]{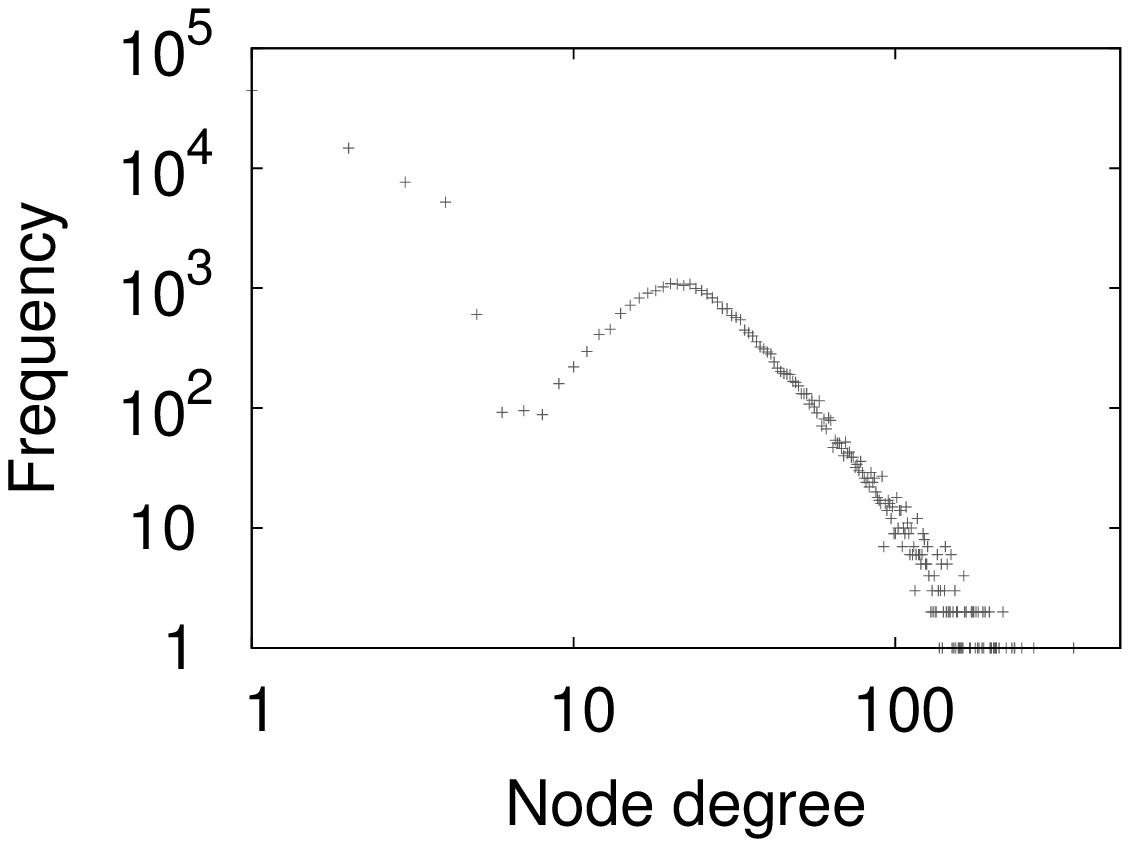}} \\
\subfloat[{IMDbNet BF}]{\label{fig:nd_imdb_0} 
\includegraphics[width=0.3\textwidth]{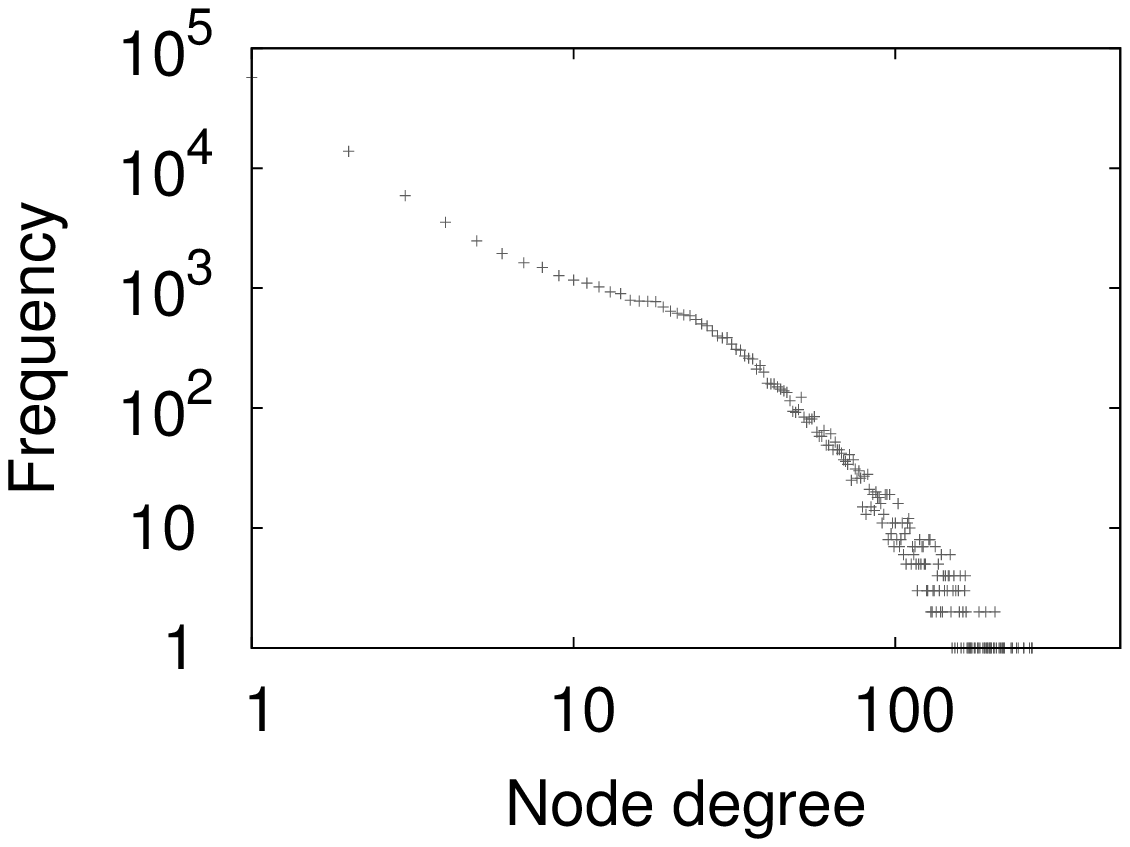}}
\subfloat[{IMDbNet $\alpha$=0.01}]{\label{fig:nd_imdb_1} 
\includegraphics[width=0.3\textwidth]{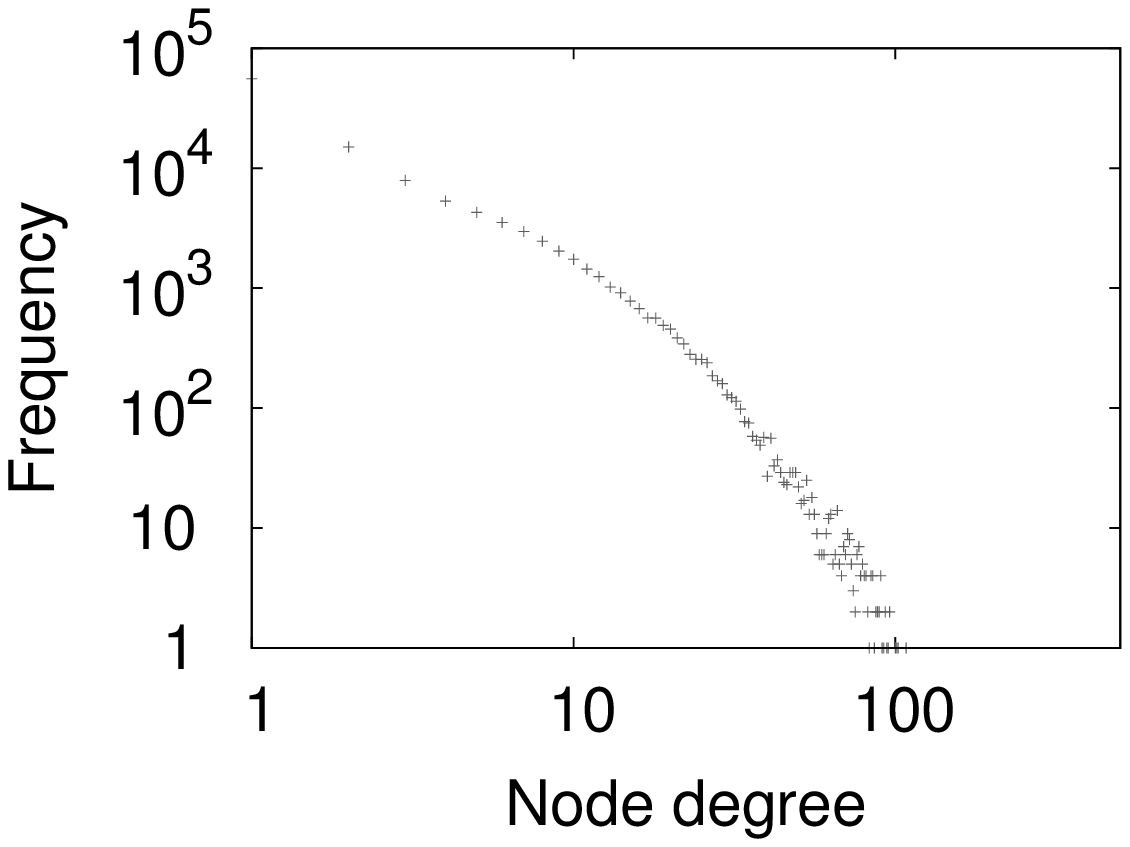}}
\subfloat[{IMDbNet $\alpha$=0.0}]{\label{fig:nd_imdb_2} 
\includegraphics[width=0.3\textwidth]{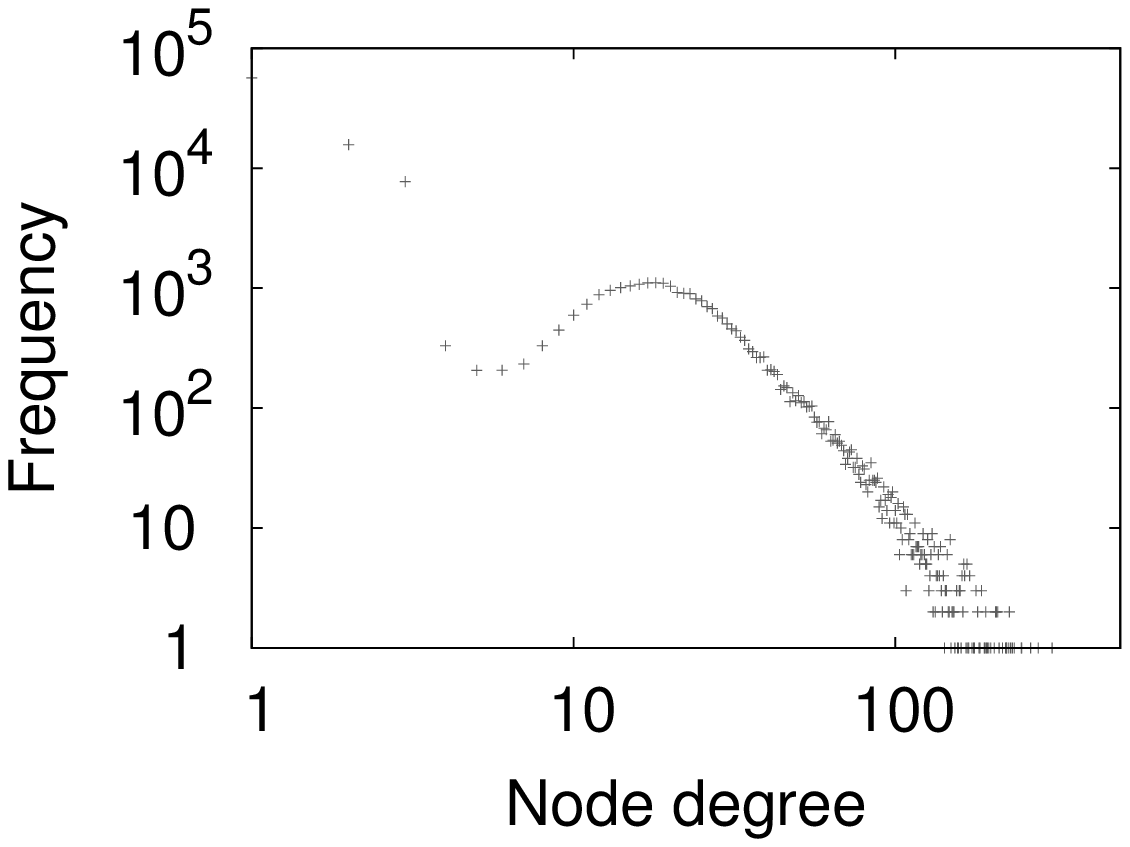}} 
\caption{Node degree distribution for different algorithms and the WikiNet/IMDbNet seed sets.} 
\label{fig:node-degrees}

\end{figure*}

Figure~\ref{fig:node-degrees} shows the degree distributions of the nodes in the constructed networks. The distribution for the breadth-first ({\em BF}) strategy is close to a power law distribution.
For the {\em Prio} strategies the described effect for the decay parameter $\alpha$ interferes
with the power law, resulting in two local extrema: a minimum for nodes with lower in-degree followed by a maximum for nodes with higher in-degree.

\begin{figure*}[!hbp]
\center
\subfloat[{WikiNet BF}]{\label{fig:nd_wikipedia_bfs}
\includegraphics[width=0.25\textwidth]{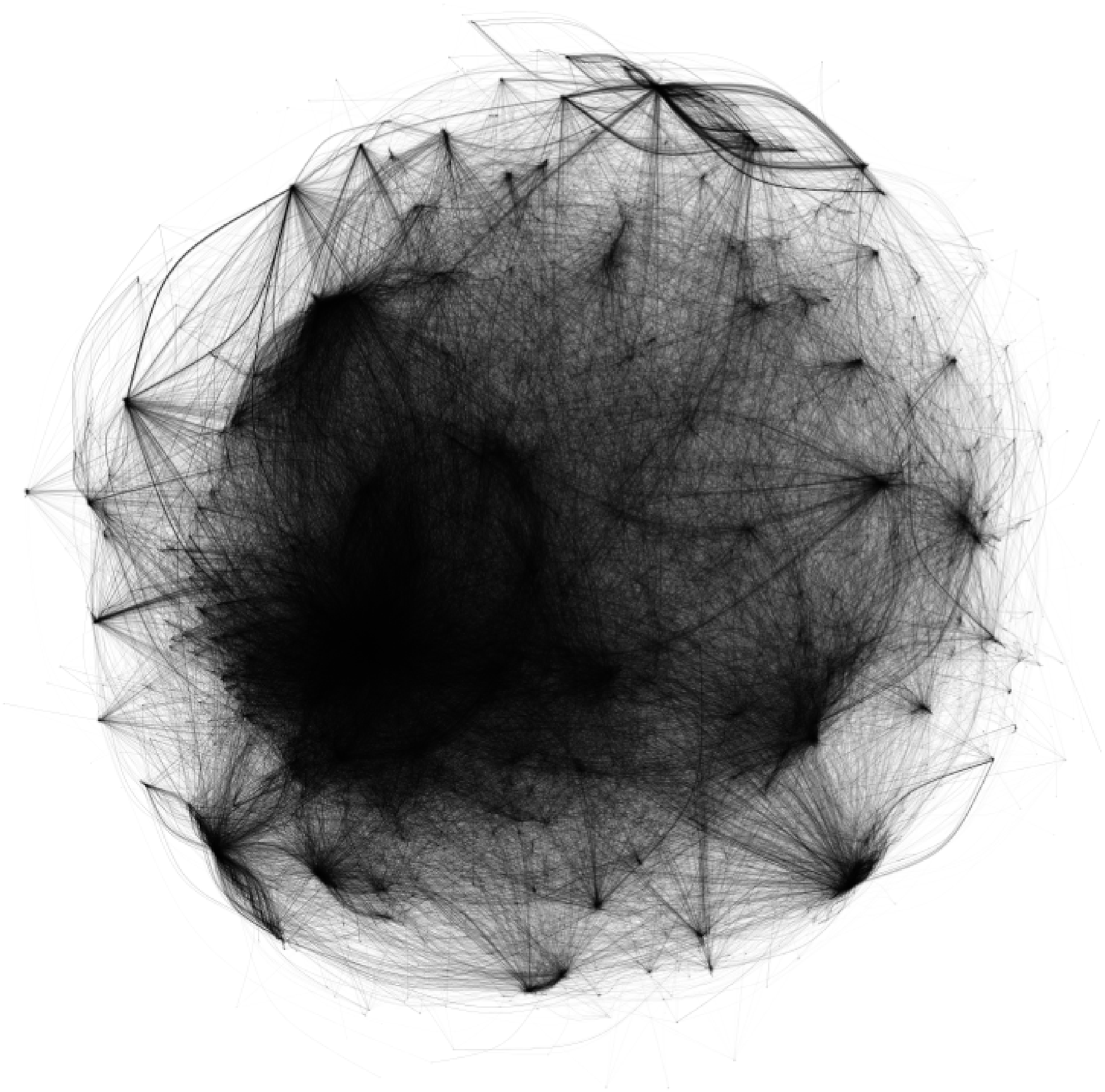}}
\subfloat[{WikiNet $\alpha$=0.0}]{
\includegraphics[width=0.25\textwidth]{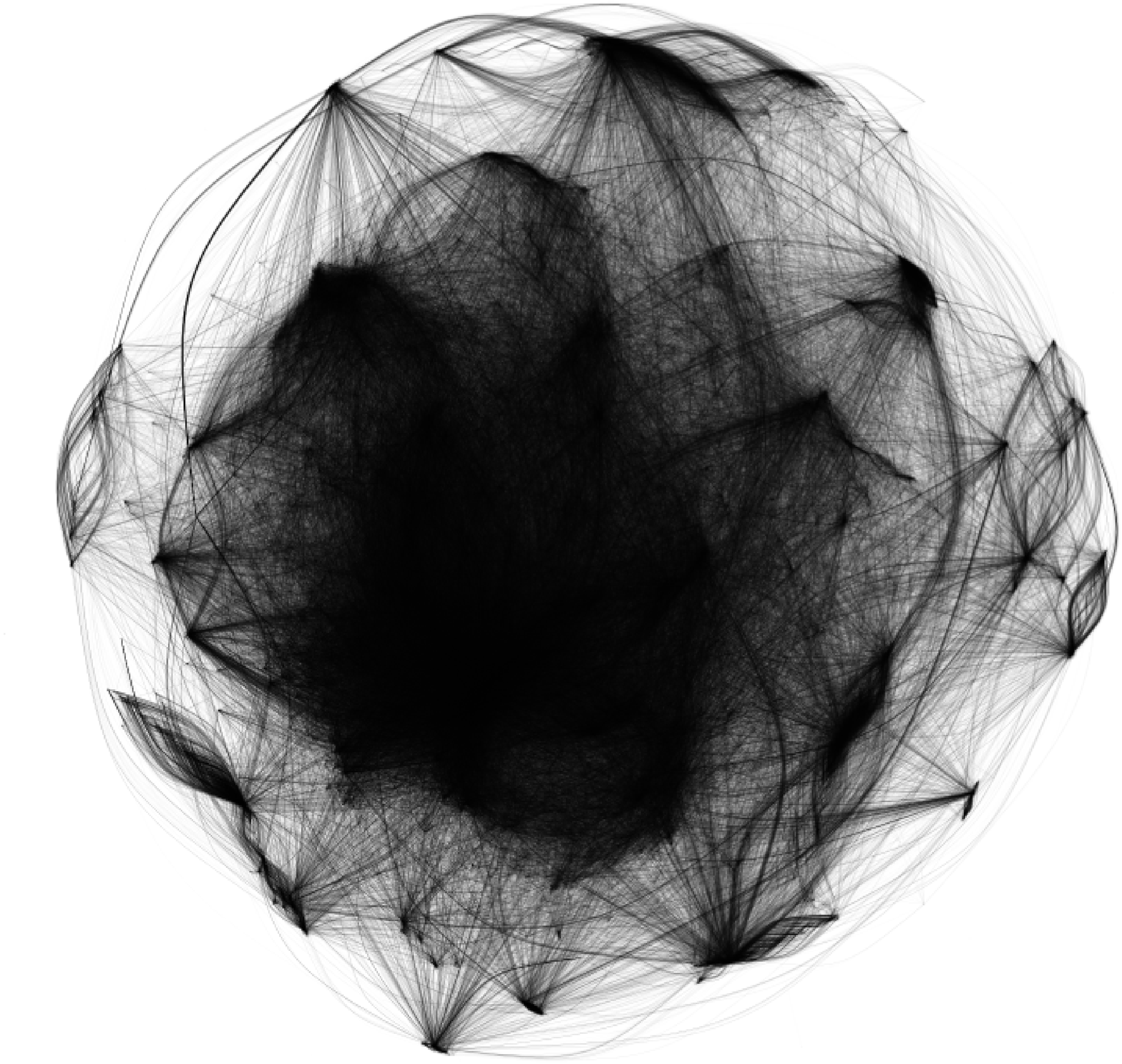}}
\subfloat[{WikiNet $\alpha$=0.01}]{\label{fig:nd_wikipedia_01}
\includegraphics[width=0.25\textwidth]{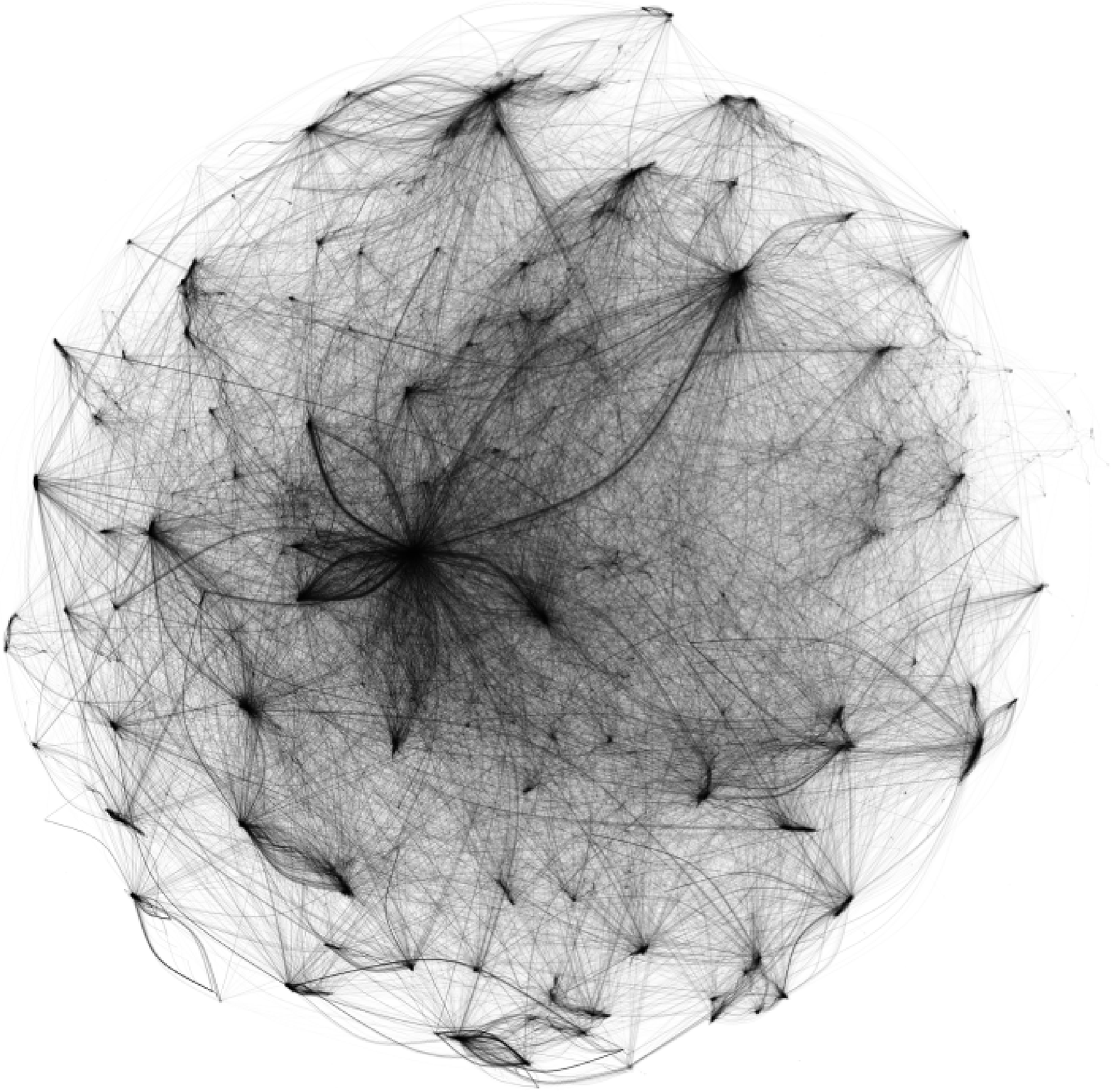}} \\
\subfloat[{IMDbNet BF}]{
\includegraphics[width=0.25\textwidth]{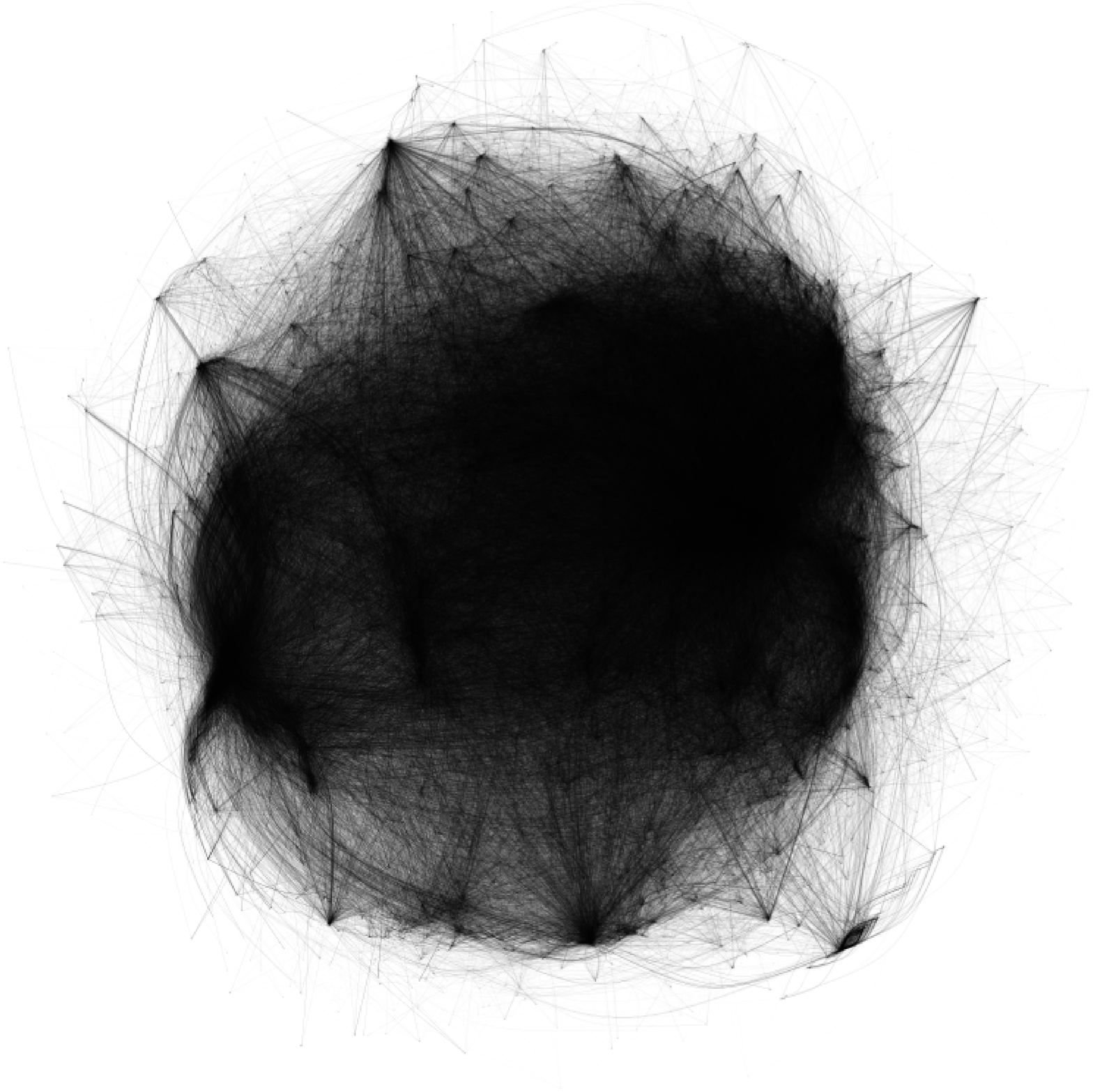}}
\subfloat[{IMDbNet $\alpha$=0.0}]{
\includegraphics[width=0.25\textwidth]{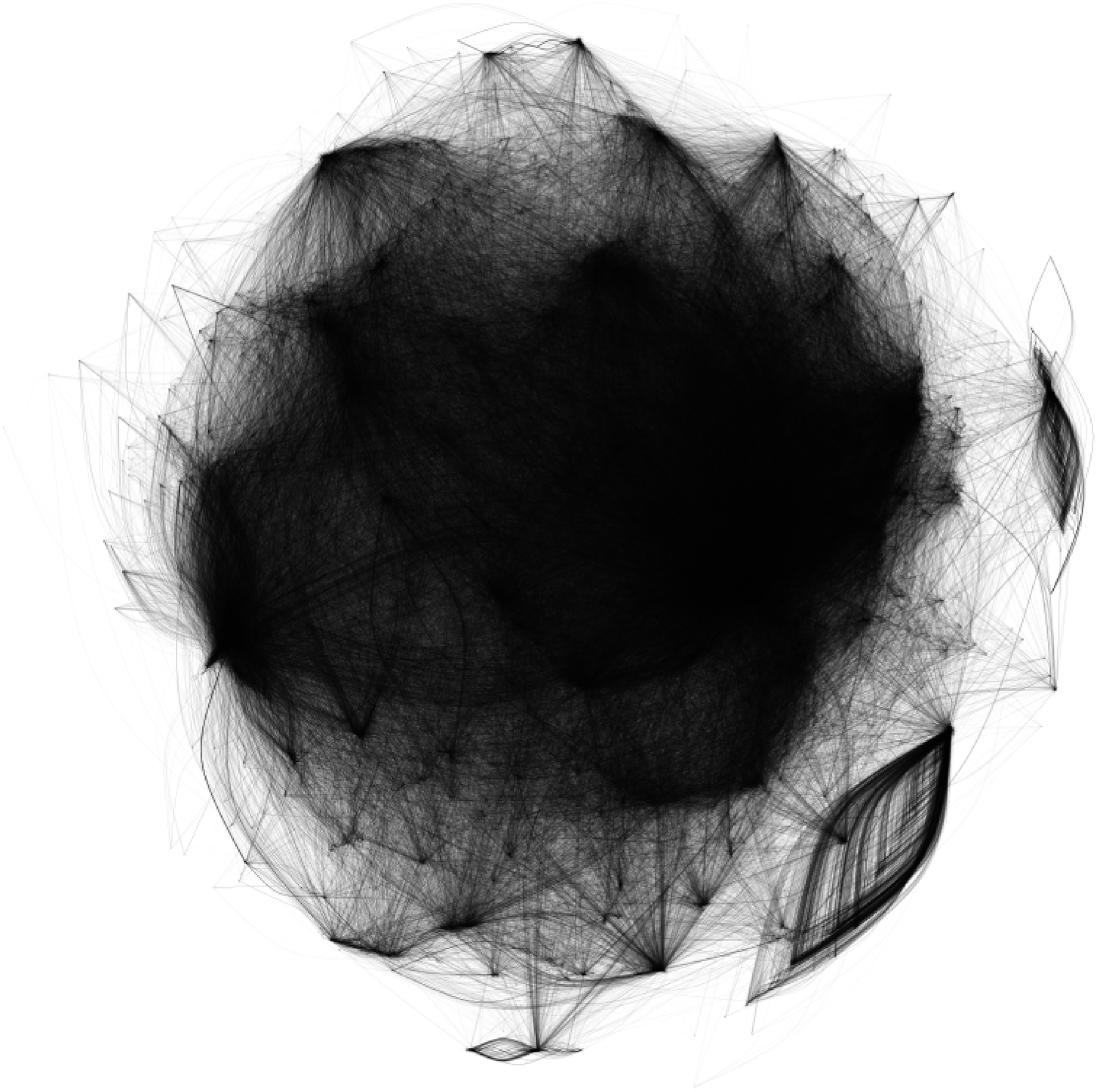}} 
\subfloat[{IMDbNet $\alpha$=0.01}]{\label{fig:nd_imdb_01} 
\includegraphics[width=0.25\textwidth]{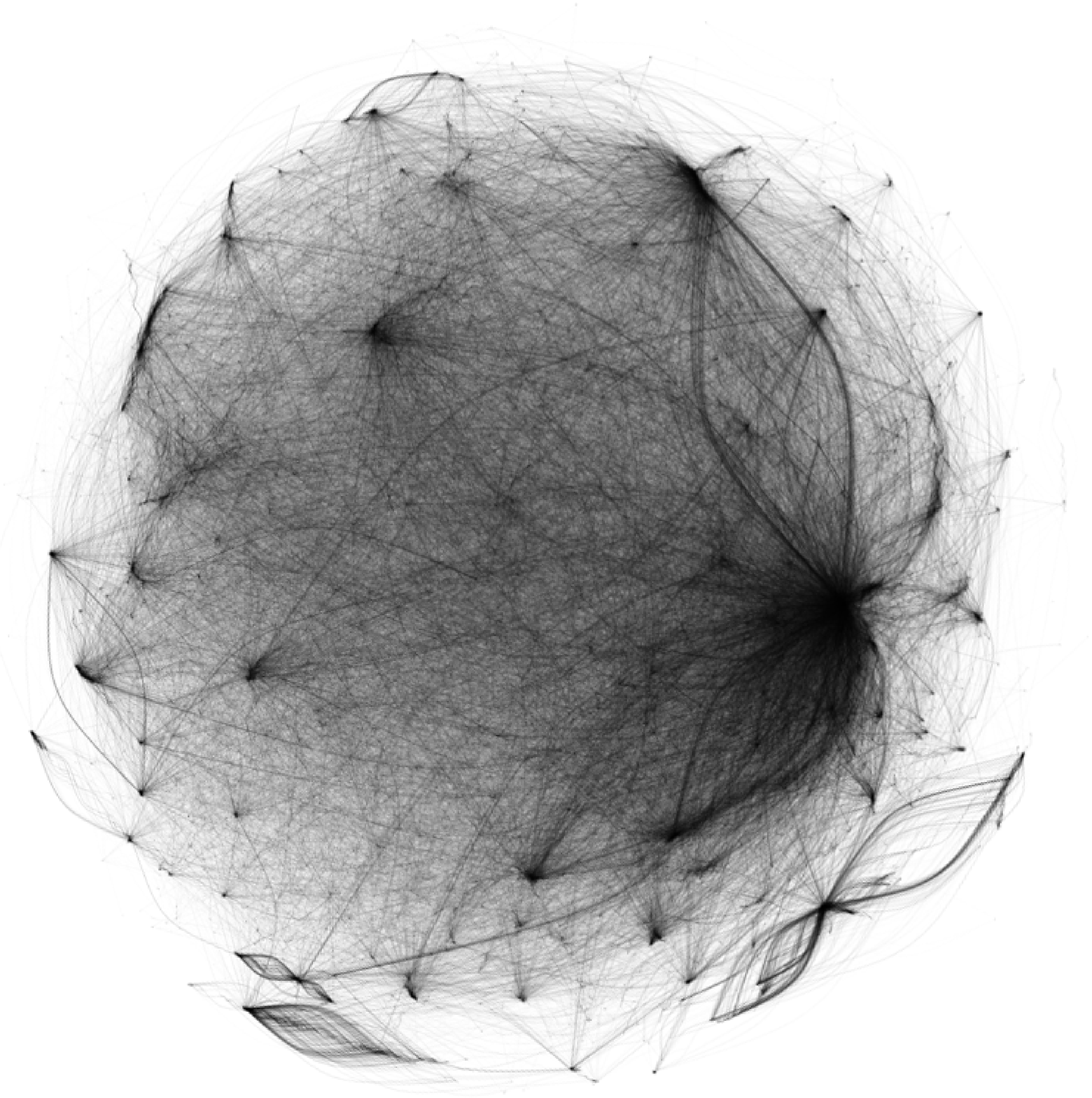}} 
\caption{Resulting WikiNet and IMDbNet graphs for different strategies and decay parameters.} 
\label{fig:hairs}
\end{figure*}

\vspace{-6pt}
\paragraph*{Graph Structures}

Figure~\ref{fig:hairs} visualizes the social networks obtained for different methods and seed sets. For the prioritization strategy {\em Prio} the decay parameter $\alpha$ has a strong influence on the network structure: 
For a stronger focus on popularity ($\alpha=0$) the larger number of edges results in higher graph density; putting more emphasis on novelty ($\alpha=0.01$) leads to a wider spread of the graph, providing a bird's eye view of the discovered space. A more in-depth exploration of community structures is subject to our future work. 

\vspace{-6pt}
\paragraph*{Example Connections}
\begin{sloppypar}
Finally, Table~\ref{tab:top-edges} lists the top-15 connections in the IMDbNet
found by the {\em BF} method. Among those connections are famous co-actors (``Terence Hill'' and ``Bud Spencer''), romantic couples (``Ryan Gosling'' and ``Eva Mendes''), or a combination of both (``Brad Pitt'' and ``Angelina Jolie'' - the strongest connection obtained). In addition, we observe that our methods are also capable of discovering connections from other domains (``Sergey Brin'' and ``Larry Page'').
\end{sloppypar}


\subsection{Quality of Extracted Networks}
\label{sec:graph_generation_quality}

We conducted a large user study to evaluate the quality of the constructed networks with overall 5,600 evaluated edges.
The goal of our evaluation was two-fold: First to check the correctness of the extracted social connections; second to evaluate strength of the connections in more detail. 
\vspace{-6pt}
\paragraph*{Pairwise assessment}
Determining the correctness of a connection between two persons is a difficult task for human assessors. Therefore, instead of letting assessors directly assign a relevance value, we first asked them to choose between two reference connections. One of the connected pairs was obtained using our method and the other consisted of persons uniformly sampled from the entity set (the order of these two options was randomized). Users were asked the following question: ``Which of the given pairs has the stronger relation?'', and were encouraged to use any source including search engines to answer this question (as an aid we provided links to a Google search page). The assessors consisted of 15 undergraduate students of Computer Science and other disciplines, and 5 graduate students of Computer Science.


\begin{sloppypar}
Table~\ref{tab:manual_accuracy} shows that all of our methods achieved accuracy values between 95\% and 99\%, demonstrating the correct extraction of almost all tested edges. Overall, for 4694 out of 4800 edge pairs the users preferred the connections discovered by our algorithms.
We measured inter-rater agreement with 5 users for a subset of 200 edges from each of the constructed networks and obtained an average pairwise percent agreement of 93.7\% and 96.5\% for WikiNet and IMDbNet, respectively, also reflected by high Fleiss' Kappa~\cite{gwet2010handbook} values of 0.83 for WikiNet and 0.97 for IMDbNet.
\end{sloppypar}
\vspace{12pt}
We also tested how many of the edges in our graphs corresponded to co-workers in movies as listed in IMDb. 
A large percentage of edges found by our algorithms (e.g. 33.8\% for method {\em BF}, and 32.7\% for method \emph{Prio} with $\alpha=0$) were co-acting in the same movies. 
However, our algorithms are capable to additionally reveal many relationships between actors not appearing in the same movie as captured by IMDb (for instance, ``Tom Cruise'' and ``Katie Holmes''). 
\begin{table}[!tp]
\setlength\tabcolsep{8pt} 
\center
\begin{tabular}{|rcl|}
\hline
\multicolumn{1}{|r}{\bf PERSON 1    } & & \multicolumn{1}{l|}{\bf PERSON 2}\\
\hline \hline
Brad Pitt    & -- & Angelina Jolie  \\
Trey Parker   & -- & Matt Stone    \\
Spencer Pratt  & -- & Heidi Montag   \\
Stephen Moyer  & -- & Anna Paquin   \\
Bud Abbott   & -- & Lou Costello   \\
Robert Pattinson & -- & Kristen Stewart \\
Bruce Robison  & -- & Kelly Willis   \\
Ryan Gosling  & -- & Eva Mendes    \\
Dev Patel    & -- & Freida Pinto   \\
Terence Hill  & -- & Bud Spencer   \\
Sergey Brin   & -- & Larry Page    \\
Nick Cannon   & -- & Mariah Carey   \\
Tod Williams  & -- & Billie Tsien   \\
Josh Dallas   & -- & Ginnifer Goodwin \\
Eric Dane    & -- & Rebecca Gayheart \\
\hline
\end{tabular}
\caption{Top-15 social relations found for IMDbNet using the BF extraction method.}
\label{tab:top-edges}
\end{table} 
\begin{table}[t!]
\center
\def\arraystretch{1.1}
\begin{tabular}{llrc|}
\cline{2-4}
\multicolumn{1}{l|}{}  & \bf Algorithm   & \bf Votes 
&\multicolumn{1}{c|}{\bf Accuracy} \\
\cline{2-4} \vspace{-3.5mm} \\
\hline
\multicolumn{1}{|l|}{\multirow{3}{*}{WikiNet}} & BF & 800 & 0.969 $\pm$
0.008  \\
\multicolumn{1}{|l|}{}      & decay a=0.0    & 800  &0.990 $\pm$
0.007  \\
\multicolumn{1}{|l|}{}      & decay a=0.01   & 800   &0.983 $\pm$
0.009  \\
\hline
\multicolumn{1}{|l|}{\multirow{3}{*}{IMDbNet}}   & BF  & 800   & 0.958  $\pm$ 0.014  \\
\multicolumn{1}{|l|}{}      & decay a=0.0    & 800   &0.983  $\pm$
0.009   \\
\multicolumn{1}{|l|}{}      & decay a=0.01   & 800   & 0.986 
$\pm$ 0.008   \\
\hline
\multicolumn{1}{|l|}{Overall}  &  & 4800   & 0.978  $\pm$ 0.004  
\\
\hline
\end{tabular}
 \vspace{-4pt}
 \caption{User evaluation results: accuracy of different\\ methods \label{tab:manual_accuracy}} 
 \vspace{-12pt}
\end{table}
\vspace{-15pt}
\paragraph*{Assessment of individual edges} 
In addition to the pairwise checks described above we also directly assessed strength and correctness of individual connections using a 5 point Likert scale. Levels 1 to 5 corresponded to the
following answers to the question ``Are these people connected?'': 1 - ``strongly disagree - Persons are not connected'', 2 - ``disagree - Persons do not likely know each other in person'', 3 - ``not
known - it is not clear whether the persons know each other in person or not'', 4 - ``agree - Persons most probably know each other in person'', 5 - ``strongly agree - Persons are strongly related or
know each other well (in person)''. We assessed 600 edges (100 edges per dataset for each of our methods) sampled from the generated social networks and mixed in random order with 200 uniformly sampled pairs of unconnected nodes. Pairs of nodes generated using our approaches obtained an average rating of 4.55 (stdv 0.87), in contrast to an average of 1.81 (stdv 0.99) for unconnected nodes. We also found that just 10.2\% of the social connections found in our networks obtained a rating of 3 or lower. These results further confirm the high accuracy of our graph construction methods.


\subsection{Pattern Expansion and Analysis}
\label{sec:pattern_analysis}
\def\arraystretch{0.9}%

\begin{table}

 \center 
\begin{tabular}{|lrrr|}
\hline
\multicolumn{1}{|c}{\bf Pattern}      & \multicolumn{1}{c}{\bf Edges} & \multicolumn{1}{c}{\bf Pairs} & \multicolumn{1}{c|}{\bf Domains} \\ 
\hline \hline
and        & 4,230 & 94  & 91   \\
,         & 1,656 & 93  & 87   \\
\&        & 828  & 90  & 95   \\
\textvisiblespace & 806  & 91  & 89   \\
und        & 1,124 & 79  & 71   \\
;         & 54  & 33  & 25   \\
:         & 36  & 27  & 24   \\
-         & 36  & 23  & 20   \\
or        & 33  & 19  & 14   \\
/         & 28  & 22  & 11   \\
with       & 20  & 15  & 15   \\
on        & 23  & 15  & 14   \\
y         & 16  & 14  & 13   \\
and his wife   & 32  & 11  & 10   \\
et        & 16  & 11  & 9    \\
|         & 18  & 13  & 7    \\
mit        & 14  & 12  & 9    \\
e         & 14  & 10  & 9    \\
vs.        & 13  & 10  & 9    \\
and wife     & 25  & 7   & 7    \\
+         & 9   & 8   & 7    \\
left and     & 13  & 7   & 6    \\
and actor     & 12  & 7   & 7    \\
hat        & 8   & 7   & 8    \\
Pictures Photo of & 31  & 28  & 1    \\ \hline
\end{tabular}
\vspace{-6pt}
\caption{Top-25 patterns after 1st iteration of pattern mining algorithm {\em PatternIter}
applied to IMDbNet}
\label{tab:Pattern_mining} 
\vspace{-12pt}
\end{table}

\begin{table*}
\centering
\renewcommand{\arraystretch}{1.25}
\small
\begin{tabular*}{0.47\textwidth}{@{\extracolsep{\fill}}|l l l l l|}
\hline
\multicolumn{5}{|c|}{\textbf{Pattern terms in WikiNet (politician seed) }} \\
\hline \hline
meet	&	minist	&	presid	&	coach	&	prime	 \\
forward	&	former	&	senat	&	rep	&	leader	 \\
center	&	beat	&	back	&	guard	&	defens	\\
king	&	gener	&	sai	&	manag	&	mayor	 \\
right	&	defend	&	sen	&	defeat	&	chief	 \\
midfield	&	lineback	&	score	&	receiv	&	captain	\\
left	&	governor	&	secretari	&	foreign	&	tackl	\\
replac	&	khan	&	player	&	against	&	face	\\
assist	&	deputi	&	premier	&	state	&	while	\\
head	&	u.	&	professor	&	republican	&	talk	\\
\hline
\end{tabular*}
\hspace{3mm}
\begin{tabular*}{0.47\textwidth}{@{\extracolsep{\fill}}|l l l l l|}
\hline
\multicolumn{5}{|c|}{\textbf{Pattern terms in IMDbNet (actor seed)}} \\
\hline \hline
star	&	direct	&	photo	&	produc	&	actor	 \\
director	&	pictur	&	wife	&	written	&	actress	 \\
galleri	&	husband	&	latest	&	daughter	&	pic	 \\
plai	&	date	&	welcom	&	film	&	cast	 \\
girlfriend	&	join	&	alongsid	&	marri	&	boyfriend	 \\
featur	&	video	&	screenwrit	&	attend	&	movi	 \\
imag	&	danc	&	perform	&	show	&	sister	 \\
camilla	&	frontman	&	screenplai	&	dure	&	skarsgard	\\
mother	&	born	&	parker	&	writer	&	comedi	\\
fan	&	also	&	stellan	&	best	&	kiss	 \\
\hline
\end{tabular*}
\caption{Top-50 stemmed terms extracted from patterns according to their Mutual Information values with seed sets of politicians in Wikipedia and actors in IMDb.} \label{tab:mitab}
\end{table*}

In this subsection we show results for the cost efficiency and accuracy of the iterative pattern mining algorithm described in Section~\ref{sec:Iterative}, and analyze the phrase patterns obtained for social connections. 
\vspace{-7pt}
\paragraph*{Cost Efficiency and Accuracy}

We conducted two iterations of the pattern mining algorithm leading to 100,319 requests for IMDbNet and 44,089 for WikiNet and obtained over 2,100 and over 5,000 distinct patterns (occurring with frequency of at least 10) for WikiNet and IMDbNet, respectively. 
The extracted patterns provide additional information about type and topical focus of the extracted social connections. 
Overall the resulting WikiNet contained 20,622 nodes and 40,342 edges, and the resulting IMDbNet 35,649 nodes and 129,326 edges.
 
A human assessment of edges extracted by the algorithm, which was conducted in the same way as described in Section~\ref{sec:graph_generation_quality}, resulted in accuracy values of more than 0.96 and 0.97 for the graph expansions of the WikiNet and IMDbNet seed sets, respectively - with 95$\%$ confidence intervals of $0.01$. This demonstrates the viability of our approach for constructing high-quality networks.  

\paragraph*{Extracted Phrase Patterns}
Table~\ref{tab:Pattern_mining}
shows the 25 best patterns extracted by searching for the 100 most highly weighted edges after the first iteration of actor seeded graph construction, and computing pattern ratings as described in Section~\ref{sec:Iterative}. 
Apart from obvious - yet useful - relationship patterns like \emph{``with''}, \emph{``or''}, and \emph{``on''}, many of the mined patterns consist of special characters such as \emph{``,''}, ``\&'', and \emph{``\textvisiblespace''} which often serve to connect persons linked to each other. Although English patterns are prevalent, our method found also useful patterns in other languages allowing for more general, multilingual retrieval of social connections, and widening the scope of the mined networks. One such pattern was, for instance, the conjunction term \emph{``and''} in Spanish, French, and German (\emph{``y''}, \emph{``et''}, and \emph{``und''}). 
Just a small number of patterns do not describe relations between persons. This holds for instance the phrase \emph{``Pictures Photos of''} which occurred in a large spam site containing many person names. However, this pattern obtained a comparably low rating because it appeared in just a single domain. Finally, we observe a few more specific patterns such as \emph{``and US president''}, \emph{``and actors''}, \emph{``and his wife''} that can help to reveal interesting information about the type of detected relationships and contexts. 

\paragraph*{Term Analysis of Patterns}
Patterns can provide useful clues on domains and topical foci of extracted (sub)networks. 
Although general patterns like \emph{``and''} and \emph{``with''} occurred, for instance, very frequently both in the context of movies and politics we also found various domain specific differences: Patterns such as \emph{ ``meets''}, \emph{``in conversation with''}, and \emph{``is joined by''} were more dominant in the politician seed based networks, while patterns like
\emph{``and director''}, \emph{``and actress''}, \emph{``is starring alongside''} were more prominent in the movie domain.

In order to obtain the most
discriminative terms from the patterns, we computed the Mutual Information (MI)
measure~\cite{Manning2008} from information theory, which measures how much the
joint distribution of features (terms from patterns in our case) and categories deviates from a
hypothetical distribution in which features and categories (``politics'' and
``movies'') are independent of each other. 
Table~\ref{tab:mitab} shows the top-50
representative \emph{stemmed} words automatically
extracted from our query patterns for the politician and actor seeded networks
described in Section~\ref{sec:graph_generation_setup}. 
Many of the terms in the patterns falling into the politics category describe personal roles like leadership 
( \emph{``minist''}, \emph{``presid''}, \emph{``senat''}) and peripheral or assisting positions ( \emph{``guard''}, \emph{``assist''}, \emph{``secretari''}), as well as
concepts related to competition (\emph{``defend''}, \emph{``score''}, \emph{``beat''}). 
In contrast, terms from patterns in the movie domain mostly refer to actors (\emph{``star''}, \emph{``actor''}, \emph{``actress''}),
film shooting (\emph{``cast''}, \emph{``screenwrit''}, \emph{``plai''}), and family relations (\emph{``husband''}, \emph{``sister''},
\emph{``boyfriend''}).
This result indicates that terms found in patterns are often strongly correlated
with social communities, and illustrates the potential merit of using machine
learning techniques for domain specific selection of pattern candidates. This opens
promising directions for focused, topic-oriented network expansion.
 

\section{Conclusions and Future Work}
\label{sec:Conclusions_and_Future_Work}

We have introduced efficient methods for extracting social networks from unstructured Web data using connectivity search queries. 
Our graph expansion algorithms leverage pattern based query mining enhanced by a bootstrapping approach for finding new query phrase patterns and by methods for intelligently prioritizing nodes. Our evaluation shows the applicability of our methods for extracting large scale social networks, and demonstrates the high accuracy of our approach. We also found that popularity and novelty of nodes are important criteria for controlling the network construction process: Popularity of nodes can be leveraged for efficiently extracting networks with more connections and higher density, while exploiting novelty is useful for faster exploration of more entities. An in-depth analysis of the characteristics of the extracted networks sheds additional light on information obtained though connectivity queries and on properties of our algorithms.

In our future work we aim to analyze pattern to reveal sentiment and other information about the social connections found: Are the linked individuals friends or foes? Do they have a personal or private connection? Are the relationships bidirectional or rather unidirectional? Furthermore, we plan to improve the efficiency of our algorithms by prioritizing search patterns based on the context of an entity using observed structures in graphs as training data for machine learning approaches. Finally, we aim to study strategies for extracting social graphs in a way that allows for simultaneously detecting communities in an efficient way. We think that this work has direct applications to social graph exploration and people search and provides a foundation for interesting analyses and findings about social networks and communities. 
\section{Acknowledgments}
This work is partly funded by the European Research
Council under ALEXANDRIA (ERC 339233) and by the
European Commission FP7 under QualiMaster (grant agreement
No. 619525).

\bibliographystyle{abbrv}
\bibliography{bibtex/stefan}

\begin{thebibliography}{10}

\bibitem{Anagnostopoulos:2008:ICS:1401890.1401897}
A.~Anagnostopoulos, R.~Kumar, and M.~Mahdian.
\newblock Influence and correlation in social networks.
\newblock KDD '08, pages 7--15. ACM, 2008.

\bibitem{bach2007review}
N.~Bach and S.~Badaskar.
\newblock {A Review of Relation Extraction}.
\newblock 2007.

\bibitem{Bird:2006:MES:1137983.1138016}
C.~Bird, A.~Gourley, P.~Devanbu, M.~Gertz, and A.~Swaminathan.
\newblock Mining email social networks.
\newblock In {\em Proceedings of the 2006 International Workshop on Mining
  Software Repositories}, MSR '06, pages 137--143. ACM, 2006.

\bibitem{Bouadjenek:2013:SNS:2484028.2484131}
M.~R. Bouadjenek, H.~Hacid, and M.~Bouzeghoub.
\newblock Sopra: A new social personalized ranking function for improving web
  search.
\newblock SIGIR '13, pages 861--864. ACM, 2013.

\bibitem{Cafarella:2009:WES:1519103.1519112}
M.~J. Cafarella, J.~Madhavan, and A.~Halevy.
\newblock Web-scale extraction of structured data.
\newblock {\em SIGMOD Rec.}, 37(4):55--61, mar 2009.

\bibitem{Canaleta:2008:SES:1566899.1566939}
X.~Canaleta, P.~Ros, A.~Vallejo, D.~Vernet, and A.~Zaballos.
\newblock A system to extract social networks based on the processing of
  information obtained from internet.
\newblock In {\em Proceedings of the 11th International Conference of the
  Catalan Association for Artificial Intelligence}, pages 283--292. IOS Press,
  2008.

\bibitem{Cimiano:2004:TSW:988672.988735}
P.~Cimiano, S.~Handschuh, and S.~Staab.
\newblock Towards the self-annotating web.
\newblock WWW '04, pages 462--471. ACM, 2004.

\bibitem{Cimiano:2005:GCC:1060745.1060796}
P.~Cimiano, G.~Ladwig, and S.~Staab.
\newblock Gimme' the context: Context-driven automatic semantic annotation with
  c-pankow.
\newblock WWW '05, pages 332--341. ACM, 2005.

\bibitem{Elson:2010:ESN:1858681.1858696}
D.~K. Elson, N.~Dames, and K.~R. McKeown.
\newblock Extracting social networks from literary fiction.
\newblock In {\em ACL'10}.

\bibitem{DBLP:conf/socinfo/WienekeDSLCLNPFTMNMHM13}
L.~W. et~al.
\newblock Building the social graph of the history of european integration -
  {A} pipeline for humanist-machine interaction in the digital humanities.
\newblock In {\em HISTOINFORMATICS}, 2013.

\bibitem{knowitall}
O.~Etzioni, M.~Cafarella, D.~Downey, S.~Kok, A.-M. Popescu, T.~Shaked,
  S.~Soderland, D.~S. Weld, and A.~Yates.
\newblock Web-scale information extraction in knowitall: (preliminary results).
\newblock In {\em WWW'09}.

\bibitem{Goyal:2010:LIP:1718487.1718518}
A.~Goyal, F.~Bonchi, and L.~V. Lakshmanan.
\newblock Learning influence probabilities in social networks.
\newblock WSDM '10, pages 241--250. ACM, 2010.

\bibitem{Guy:2010:SMR:1835449.1835484}
I.~Guy, N.~Zwerdling, I.~Ronen, D.~Carmel, and E.~Uziel.
\newblock Social media recommendation based on people and tags.
\newblock SIGIR '10, pages 194--201. ACM, 2010.

\bibitem{gwet2010handbook}
K.~Gwet.
\newblock {\em Handbook of Inter-rater Reliability: The Definitive Guide to
  Measuring the Extent of Agreement Among Raters}.
\newblock Advanced Analytics, LLC, 2010.

\bibitem{he2007efficient}
J.~He, Y.~Liu, Q.~Tu, C.~Yao, and N.~Di.
\newblock Efficient entity relation discovery on web.
\newblock {\em JCIS}, 2007.

\bibitem{Jain:2009:ICE:1645953.1646198}
A.~Jain and P.~Pantel.
\newblock Identifying comparable entities on the web.
\newblock CIKM '09, pages 1661--1664. ACM, 2009.

\bibitem{Jiang:2013:LOC:2541176.2505666}
Z.~Jiang, L.~Ji, J.~Zhang, J.~Yan, P.~Guo, and N.~Liu.
\newblock Learning open-domain comparable entity graphs from user search
  queries.
\newblock CIKM, 2013.

\bibitem{Jin:2010:TID:1772690.1772740}
X.~Jin, S.~Spangler, R.~Ma, and J.~Han.
\newblock Topic initiator detection on the world wide web.
\newblock WWW '10, pages 481--490. ACM, 2010.

\bibitem{ottoman}
C.~Karbeyaz, E.~Can, F.~Can, and M.~Kalpakli.
\newblock A content-based social network study of evliya celebi's
  seyahatname-bitlis section.
\newblock In {\em Computer and Information Sciences II}. Springer, 2012.

\bibitem{DBLP:journals/aim/KautzSS97}
H.~A. Kautz, B.~Selman, and M.~A. Shah.
\newblock The hidden web.
\newblock {\em {AI} Magazine}, 18(2):27--36, 1997.

\bibitem{Kleinberg:1999:HAC:345966.345982}
J.~M. Kleinberg.
\newblock Hubs, authorities, and communities.
\newblock {\em ACM Comput. Surv.}, 31(4es), dec 1999.

\bibitem{konstant2014}
N.~Konstantinova.
\newblock Review of relation extraction methods: What is new out there?
\newblock Springer, 2014.

\bibitem{Langville:2006:GPB:1146372}
A.~N. Langville and C.~D. Meyer.
\newblock {\em Google's PageRank and Beyond: The Science of Search Engine
  Rankings}.
\newblock Princeton University Press, 2006.

\bibitem{Manning2008}
C.~Manning, P.~Raghavan, H.~Schutze, and E.~Corporation.
\newblock {\em {Introduction to information retrieval}}.
\newblock Cambridge University Press, 2008.

\bibitem{Matsuo:2006:PAS:1135777.1135837}
Y.~Matsuo, J.~Mori, M.~Hamasaki, K.~Ishida, T.~Nishimura, H.~Takeda, K.~Hasida,
  and M.~Ishizuka.
\newblock Polyphonet: An advanced social network extraction system from the
  web.
\newblock WWW, 2006.

\bibitem{DBLP:conf/aaai/MatsuoTN07}
Y.~Matsuo, H.~Tomobe, and T.~Nishimura.
\newblock Robust estimation of google counts for social network extraction.
\newblock In {\em AAAI}, 2007.

\bibitem{Mika:2005:FSW:1741305.1741326}
P.~Mika.
\newblock Flink: Semantic web technology for the extraction and analysis of
  social networks.
\newblock {\em Web Semant.}, 3(2-3):211--223, oct 2005.

\bibitem{Mohaisen:2010:MMT:1879141.1879191}
A.~Mohaisen, A.~Yun, and Y.~Kim.
\newblock Measuring the mixing time of social graphs.
\newblock In {\em SIGCOMM}, 2010.

\bibitem{DBLP:conf/rskt/NasutionN10}
M.~K.~M. Nasution and S.~A. Noah.
\newblock Superficial method for extracting social network for academics using
  web snippets.
\newblock In {\em RSKT}, 2010.

\bibitem{nuray2009exploiting}
R.~Nuray-Turan, Z.~Chen, D.~V. Kalashnikov, and S.~Mehrotra.
\newblock Exploiting web querying for web people search in weps2.
\newblock In {\em Web People Search Evaluation Workshop (WePS), 18th WWW
  Conference}, 2009.

\bibitem{ilprints422}
L.~Page, S.~Brin, R.~Motwani, and T.~Winograd.
\newblock The pagerank citation ranking: Bringing order to the web.
\newblock Technical report, 1999.

\bibitem{DBLP:conf/acl/PantelP06}
P.~Pantel and M.~Pennacchiotti.
\newblock Espresso: Leveraging generic patterns for automatically harvesting
  semantic relations.
\newblock In {\em {ACL} 2006, Sydney, Australia}, 2006.

\bibitem{Pasca:2004:ACN:1031171.1031194}
M.~Pasca.
\newblock Acquisition of categorized named entities for web search.
\newblock In {\em CIKM'04}.

\bibitem{Tang:2008:AEM:1401890.1402008}
J.~Tang, J.~Zhang, L.~Yao, J.~Li, L.~Zhang, and Z.~Su.
\newblock Arnetminer: Extraction and mining of academic social networks.
\newblock KDD '08, pages 990--998. ACM, 2008.

\bibitem{Yates:2007:TOI:1614164.1614177}
A.~Yates, M.~Cafarella, M.~Banko, O.~Etzioni, M.~Broadhead, and S.~Soderland.
\newblock Textrunner: Open information extraction on the web.
\newblock In {\em ACL}, 2007.

\bibitem{SocialInfluencePropagation}
S.~Ye and S.~Wu.
\newblock Measuring message propagation and social influence on twitter.com.
\newblock In {\em LNCS}. 2010.

\bibitem{Yu:2004:TDS:1013367.1013519}
P.~S. Yu, X.~Li, and B.~Liu.
\newblock On the temporal dimension of search.
\newblock WWW, 2004.

\end{thebibliography}
\balance

\end{document}